\documentclass[pra,aps,twocolumn,10pt,floatfix,longbibliography,superscriptaddress]{revtex4-2}

\usepackage[usenames,dvipsnames]{xcolor}
\usepackage{soul}
\usepackage{amsmath}
\usepackage{amsfonts}
\usepackage{txfonts}
\usepackage{amssymb}
\usepackage{xr-hyper}
\usepackage[colorlinks=true,citecolor=cyan,linkcolor=magenta,filecolor=magenta]{hyperref}
\usepackage[capitalize]{cleveref}
\usepackage{graphicx}
\usepackage{bbold}					
\usepackage[makeroom]{cancel}		
\usepackage{multirow}				
\usepackage[normalem]{ulem}         
\usepackage{array}
\usepackage{mathrsfs}
\usepackage{mathtools}
\usepackage[version=4]{mhchem}      
\usepackage{IEEEtrantools}          
\usepackage{dcolumn}                
\usepackage{physics}                
\usepackage{dsfont}                 
\usepackage{enumerate}              
\usepackage[shortlabels]{enumitem}  
\usepackage{verbatim}
\usepackage{csquotes}
\usepackage{pifont}

\usepackage{mathdots}
\graphicspath{{./img/}}

\newcommand{\beq}[1]{\begin{equation}\label{#1}}
\newcommand{\eep}{\;.\end{equation}}
\newcommand{\eec}{\;,\end{equation}}
\newcommand{\eeq}{\end{equation}}

\newcommand{\la}{\lambda}

\newcommand{\sect}[1]{\vspace{0.3em}{\it #1.}---}
\DeclareMathAlphabet{\mathcal}{OMS}{cmsy}{m}{n} 

\newcommand{\black}[1]{\textcolor{black}{#1}}

\renewcommand{\vec}[1]{{\bf #1}}

\newcommand{\kv}{\vec{k}}

\newcommand{\A}{\vec{A}}

\definecolor{AB}{rgb}{1.0, 0.11, 0.81}

\begin{document}


\title{Disorder-induced topological quantum phase transitions in \black{multi-gap} Euler semimetals}

\author{Wojciech J. Jankowski}
\email{wjj25@cam.ac.uk}
\affiliation{TCM Group, Cavendish Laboratory, Department of Physics, J J Thomson Avenue, Cambridge CB3 0HE, United Kingdom}

\author{Mohammadreza Noormandipour}
 \affiliation{TCM Group, Cavendish Laboratory, Department of Physics, J J Thomson Avenue, Cambridge CB3 0HE, United Kingdom}

\author{Adrien Bouhon}
\affiliation{TCM Group, Cavendish Laboratory, Department of Physics, J J Thomson Avenue, Cambridge CB3 0HE, United Kingdom}
\affiliation{Nordita, Stockholm University and KTH Royal Institute of Technology, Hannes Alfv{\'e}ns v{\"a}g 12, SE-106 91 Stockholm, Sweden}
 
\author{Robert-Jan Slager}
\email{rjs269@cam.ac.uk}
\affiliation{TCM Group, Cavendish Laboratory, Department of Physics, J J Thomson Avenue, Cambridge CB3 0HE, United Kingdom}

\date{\today}

\begin{abstract}
We study the effect of disorder in systems having a non-trivial Euler class. As these recently proposed multi-gap topological phases come about by braiding non-Abelian charged band nodes residing between different bands to induce stable pairs within isolated band subspaces, novel properties may be expected. Namely, a~modified stability and critical phases under the unbraiding to metals can arise, when the disorder preserves the underlying $C_2\cal{T}$ or $\cal{P}\cal{T}$ symmetry on average. Employing elaborate numerical computations, we verify the robustness of associated topology by evaluating the changes in the average densities of states and conductivities for different types of disorders. Upon performing a scaling analysis around the corresponding quantum critical points we retrieve a universality for the localization length exponent of $\nu = 1.4 \pm 0.1$ for Euler-protected phases, relating to two-dimensional percolation models. We generically find that quenched disorder drives Euler semimetals into critical metallic phases. Finally, we show that magnetic disorder can also induce topological transitions to quantum anomalous Hall plaquettes with local Chern numbers determined by the initial value of the Euler invariant. 
\end{abstract}

\pacs{Valid PACS appear here}

\maketitle


\sect{Introduction}
The study of topological semimetals has become a prominent forefront of condensed matter physics over the past decade~\cite{Armitage_2018}. With the surge of interest in graphene \cite{doi:10.1126/science.1102896, Geim2007-vq, RevModPhys.81.109, RevModPhys.83.407} due to its readily experimental availability, a focus proceeded towards three-dimensional (3D) Dirac and Weyl semimetals \cite{Chiu_2016, doi:10.1146/annurev-conmatphys-031016-025225, doi:10.1126/science.aaf5037, Volovik_2017, Armitage_2018,volovik2018investigation}, the pursuit of which by now has found experimental realizations in materials such as TaAs, NbP and TaP~\cite{doi:10.1126/science.aaa9297, PhysRevX.5.031013, Shekhar2015, doi:10.1126/science.aad8766}. Arguably, these pursuits have also fueled the intensive research efforts in topological materials as the presence of an unpaired Dirac fermion on the boundary is one of the most notable consequences of many band topologies~\cite{Rmp1,Rmp2}.

\begin{figure}[ht]
\includegraphics[width=0.95\columnwidth]{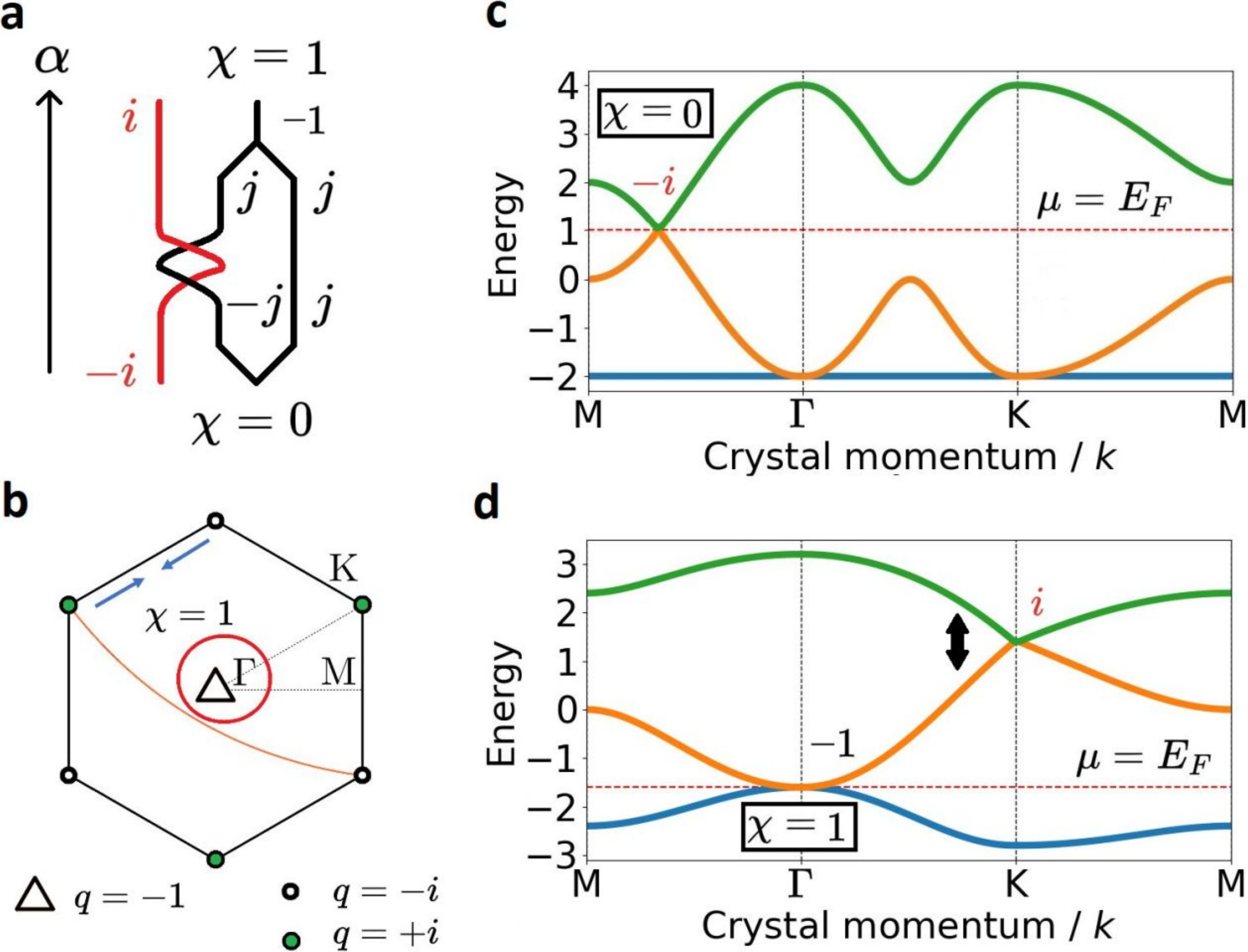}
\caption{Momentum space structure of Euler semimetals. $\textbf{(a)}$~Braiding nodes residing between different gaps (denoted by charges $i$ and $j$) results in a two-band subspace with similarly-valued nodes as function of 'time parameter' $\alpha$. A stable pair of nodes with non-zero patch Euler class and quadratic dispersion can for example be obtained by merging two nodes of identical frame charge $q$. $\textbf{(b)}$ Euler nodes in the Brillouin zone.  Blue arrows denote nodes that can be annihilated. The line indicates a Dirac string that when crossed changes the frame charge $q$ to $-q$. The red circle indicates a patch with Euler class $\chi=1$ hosting a stable quadratic node. $\textbf{(c)}$ Sections of a semimetallic band structure (from Eq.~\eqref{eq:kagomeH} with $(t,t') = (0,1)$) occupied up to the nodes (dotted red line) without Euler class and protection, along high-symmetry lines from $\textbf{(b)}$. $\textbf{(d)}$ An Euler semimetal with $(t,t') = (1,-0.2)$ protected by the patch Euler class $\chi=1$ of the two superposed nodes at $\Gamma$. Importantly, two bands (black arrow) with $\chi = 0$ can be gapped. Such gap prevents braiding which can remove the protection.}
\label{fig:fig1}
\end{figure}

As defects and disorder are realistically unavoidable in any physical realization, their fundamental role and interplay with the present topology are of primary importance~\cite{Dima1983}. In Weyl semimetals for example, quenched disorder was found to induce quantum phase transitions (QPTs) to metals~\cite{PhysRevB.96.201401, PhysRevX.8.031076, PhysRevB.93.201302, PhysRevX.6.021042,pixleyreview}, further succeeded by strong localization transitions to Anderson insulators \cite{Phil_G4} for larger disorder strengths. In addition, deeper connections between general percolation arguments as captured by the Chalker-Coddington model~\cite{JTChalker_1988} and paradigmatic Chern insulators have been established~\cite{PhysRevLett.61.2015,Song_2021}.

Given this importance we address here the effects of disorder in systems with finite Euler class~\cite{RJ_ZrTe,PhysRevX.9.021013}. This recently proposed invariant thrives on the concept of multi-gap topology~\cite{PhysRevB.102.115135}, going beyond well established symmetry-indicated paradigms~\cite{fukane, prx_us, Po2017,Slager_NatPhys_2013, Codefects2, Bradlyn:2017}. Indeed, rather than considering how bands transform at high symmetry points and induce relations between irreducible representations in which they transform, these phases arise by momentum space braiding of band nodes residing between different bands. For systems having $C_2\cal{T}$ (two-fold rotations and time-reversal) or spinless $\cal{P}\cal{T}$ (parity and time-reversal) symmetry the Hamiltonian can always be brought in real form~\cite{RJ_ZrTe,PhysRevX.9.021013}. Accordingly, band nodes between different bands can be shown by homotopy arguments~\cite{PhysRevB.102.115135} to have non-Abelian frame charges, akin to $\pi$-vortices in biaxial nematics \cite{RevModPhys.84.497, volovik2018investigation,PhysRevX.6.041025}, coinciding with the quaternion charges $\mathbb{Q} = \{\pm 1, \pm i, \pm j, \pm k \}$ for three-band systems and the so-called Salingaros group for many-band systems~\cite{doi:10.1126/science.aau8740, RJ_ZrTe}. Physically, the frame charges of these nodes correspond to different accumulated angles acquired by the eigenvector frame when parallel-transported over loops around nodes corresponding to points of band-energy-crossing between adjacent bands~\cite{RJ_ZrTe,PhysRevX.9.021013}. As a result, braiding nodes in momentum space can result in band subspaces with similarly-valued nodes that can no longer be annihilated, see Fig.~\ref{fig:fig1}, in contrast to pairs of Weyl nodes that act as monopole sources and sinks of Abelian Berry curvature that can be gapped upon bringing them together. The stability of the nodes within this two-band subspace is quantified by the Euler class invariant $\chi$ over any patch $\mathcal{D}$ in the Brillouin zone (BZ) that excludes nodes from other band spaces~\cite{RJ_ZrTe},
\begin{equation}
  \chi = \frac{1}{2\pi} \int_{\mathcal{D} \in \text{BZ}} \dd ^{2} \textbf{k}~\text{Eu} - \frac{1}{2\pi} \oint_{\partial\mathcal{D}} \dd \textbf{k} \cdot \vec{a},
  \label{eq:eulerpatch}
\end{equation}
where $\vec{a}$ is the Euler connection defined in terms of the Pfaffian of the non-Abelian Berry connection $\A_{n,n+1}(\kv) = \bra{u_n}\ket{\nabla_{\kv} u_{n+1}}$ and $\text{Eu} = \nabla_{\textbf{k}} \cross \vec{a}$ is the Euler curvature, both defined from two successive real Bloch eigenvectors $\{\vert u_{n}\rangle,\vert u_{n+1}\rangle\}$. Considering more bands and different partitions thereof, more general multi-gap phases and interplays with crystalline symmetry can arise 
to induce Euler-characterized band subspaces that can be gapped or connected to the other parts of the spectrum, characterized by richer homotopies on the associated Grassmannian or flag classifying spaces~\cite{PhysRevB.102.115135}.
Most importantly, however, these phases are increasingly being related to novel physical effects in and out of equilibrium in systems ranging from metamaterials and trapped-ion insulators to electronic and phononic spectra in both experimental and theoretical settings~\cite{PhysRevB.102.115135, RJ_meron,RJ_Kagome,RJ_ZrTe,doi:10.1126/science.aau8740, PhysRevB.100.195135, PhysRevB.105.214108, Peng2021, PhysRevB.105.L081117,bouhon2023quantum, PhysRevX.9.021013,PhysRevLett.126.246801,slager2022floquet, Jiang1Dexp, _nal_2020,Guo1Dexp,ahnprl}.

In this work, we focus on the effects of disorder in Euler semimetals in two spatial dimensions. As the non-triviality of Euler class provides an obstruction to annihilating the associated nodes, one may expect an extended regime of disorder, rather than a transition point, to 'unbraid' the nodes and induce a bulk delocalization transition to a metal. To study such QPTs quantitatively, we perform a scaling analysis at corresponding quantum critical points found numerically from average density of states (ADOS) calculations using the kernel polynomial method (KPM) \cite{KPM2006}. We deduce associated universal dynamical scaling and localization length exponents, consistent with classical 2D percolation and quantum network models~\cite{Song_2021}. Moreover, in meronic Euler phases~\cite{RJ_Kagome}, that is three-band phases with odd $\chi$, we analyze the effects of disorder on edge modes due to a $\pi$-Zak phase, which we find to be protected by braiding up to the critical points, contrary to the Fermi arcs of Weyl semimetals dissolving at subcritical disorder~\cite{PhysRevB.96.201401,pixleyreview}. Finally, on adding magnetic disorder, breaking time-reversal ($\mathcal{T}$) symmetry locally, as well as $\mathcal{C}_2{\mathcal{T}}$ or $\mathcal{PT}$ protecting the Euler invariant, while preserving these symmetries on average~\cite{PhysRevX.13.031016}, we show an emergence of plaquettes with local Chern numbers descendent from the Euler class. This corroborates with the observation in Ref.~\cite{PhysRevB.102.115135} that adding such effectively generated $\cal{T}$-breaking terms already gaps the nodes into Chernful bands in the non-disordered case.

\sect{Model setting}
For concreteness we analyze three-band and four-band Euler semimetals (see App.~\ref{app:appA}) subject to disorder. The three-band cases can in particular be modeled  on a kagome lattice (see Fig.~5, App.~\ref{app:appA}), which will be the focus of this work although we have checked our findings also in the contexts of other lattices, such as the square models of Refs.~\cite{PhysRevB.102.115135,multigap}. To generate nodes with a patch Euler class $\chi = 1$ we require only nearest-neighbor $(t)$ and next-nearest-neighbor hoppings $(t')$, originally setting the onsite energies of all orbitals to $\varepsilon_i = 0$. The corresponding tight-binding Hamiltonian can be written as
\\
\begin{equation}
\label{eq:kagomeH}
    H = \sum_i \varepsilon_i c^\dagger_i c_i + t \sum_{\langle ij \rangle}(c^\dagger_i c_j + \textnormal{h.c.}) 
    + t' \sum_{\langle\langle ij \rangle\rangle}(c^\dagger_i c_j + \textnormal{h.c.}).
\end{equation}
The patch Euler class ensures topological protection of a pair of nodes, as long as a multi-gap node braiding (Fig.~\ref{fig:fig1}(a)) does not trivialize it to $\chi = 0$ in the semimetallic subspace. Such trivialization can be prevented by gapping out pairs of oppositely charged nodes, which, as we show, ensures the protection up to closing this gap with disorder. The gap closing is necessary for the re-creation of the quaternion-valued nodes between the upper bands, as these can unbraid with the nodes residing between the bottom bands, trivializing the invariant. For the definitions of the Euler invariant and unbraiding in weakly-disordered systems lacking translational symmetries, see the App.~\ref{app:appB}. Once the upper gap closes at the critical point, the unbraiding accompanied by a bulk delocalization transition to a metal occurs, which is studied next. As long as the nodes are protected, we observe a stability of an associated edge mode in a continuous window, which we refer to as an \textit{unbraiding} regime, as detailed in the subsequent sections. We stress that edge states in Euler phases have been corroborated in this and several other models using either a Zak phase argument~\cite{Peng2021,RJ_Kagome} or an intricate interplay between Wyckoff-position and twisted-band induced Zak phases~\cite{RJ_meron} but that a full bulk-boundary correspondence for Euler class is still an open question. For our purposes, it suffices that there exists an edge state in the discussed $\chi=1$ configuration in the clean limit~\cite{Peng2021,RJ_meron,RJ_Kagome, Guo1Dexp}. We then subsequently analyze its behavior subject to disorder. 

\sect{Disorder-induced QPT of Euler semimetal-metals}
We now characterize the criticality of the Euler semimetal-metal QPTs under disorder by deducing dynamical scaling exponents $(z)$ and localization exponents $(\nu)$ (see App.~\ref{app:appC}). In three-band kagome models with nodes of {\it opposite}, non-stable, quaternion charges ($q=\pm i$), such as $(t, t') = (0, 1)$ (Fig.~\ref{fig:fig1}(c)), we find $z = 0.9 \pm 0.1$ and $\nu = 1.0 \pm 0.1$ (see App.~\ref{app:appD}), as in graphene \cite{PhysRevLett.118.036601, PhysRevB.95.075116}, consistently with no braiding-based topological protection against disorder.
\begin{figure}[h]
\includegraphics[width=\columnwidth]{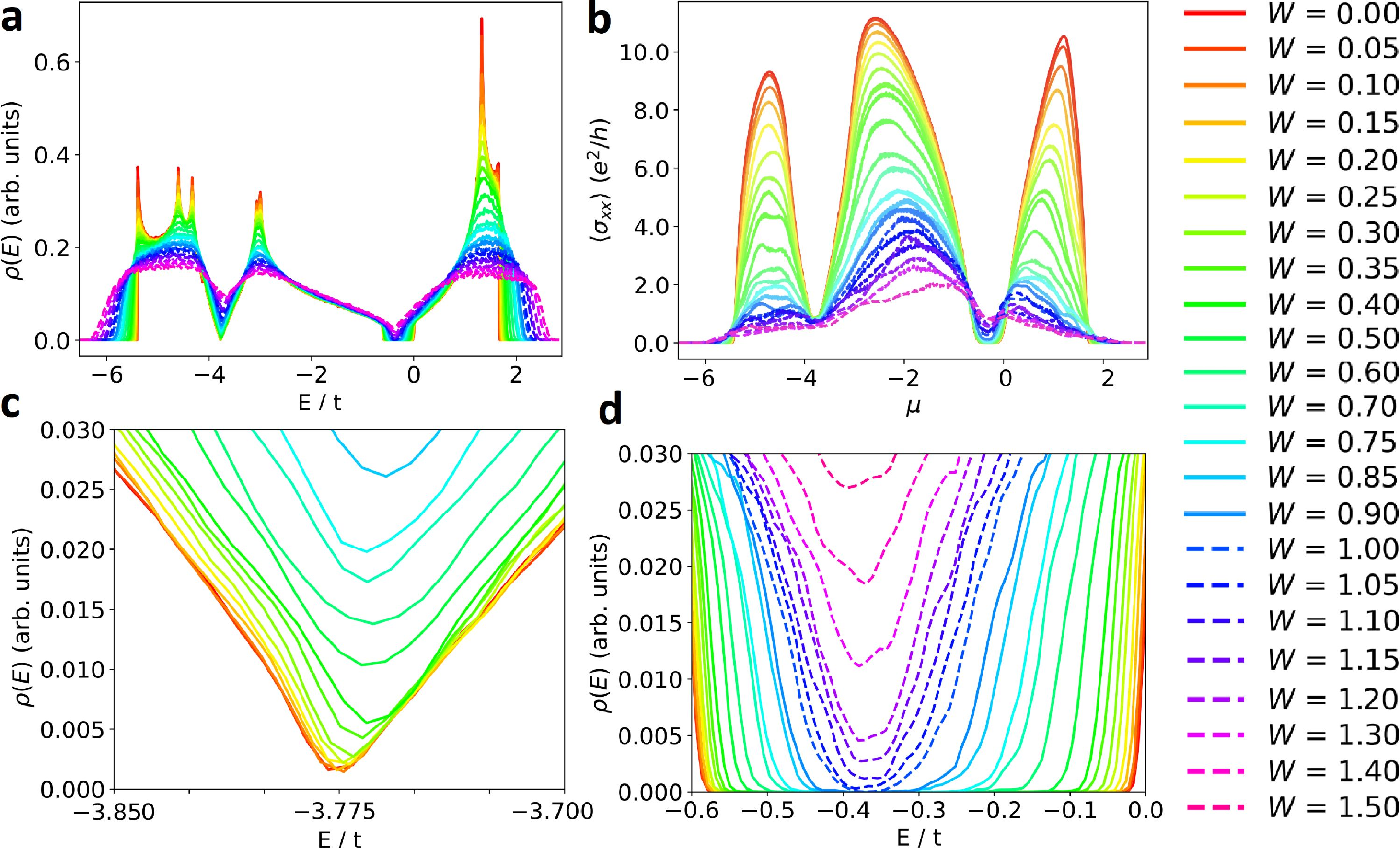}
\caption{Scaling of the ADOS $\rho(E)$ \textbf{(a)} and averaged conductivity~\textbf{(b)} up to bulk criticality. ADOS scaling around the Euler nodes  \textbf{(c)} and at gap-closing quantum critical point ($W_c = 1.05 \pm 0.05$) \textbf{(d)} of braided three-band Euler semimetal with $\chi = 1$. The scaling of conductivity and $\rho(E)$ vs. $E$ determines the dynamical scaling exponent $z$ (see also App.~\ref{app:appC}); $\rho(0)$ vs. $(W-W_c)/W_c$ describes the critical scaling of localization length exponent, which we find universally to be ${\nu = 1.4 \pm 0.1}$ for non-trivial Euler class.}
\label{fig:fig2}
\end{figure}

Intriguingly, we can contrast this result and consider a filling up till nodes having a non-trivial invariant $\chi$. To this end we consider a protected, stable, Euler semimetal by taking $(t, t') = (1, -0.2)$, which isolates Euler nodes around the BZ $\Gamma$ point from the flat band, see Fig.~\ref{fig:fig1}(d), and gapping out the top two bands by annihilating nodes with frame charges $q=\pm i$, see Fig.~\ref{fig:fig1}(b). The latter can be achieved by adding a mass term of the form $\text{diag}(-1, -1, 0)$, which breaks the $C_6$-symmetry of the kagome lattice, symmetrically splitting the $\Gamma$ node into two nodes of the same quaternion charge, maintaining an Euler patch around $\Gamma$ as in Fig.~\ref{fig:fig1}(b). Taking a filling up to these nodes results in a stable Euler semimetal. That is, the gap to the top band protects the $\chi$ invariant of the bottom two bands from trivializing by unbraiding. We can subsequently study how disorder then induces a QPT upon closing the gap and facilitating this process.

At criticality corresponding to closing a gap (Fig.~\ref{fig:fig2}), we universally find $z = 0.7 \pm 0.1$ and $\nu = 1.4 \pm 0.1$, with the localization length scaling exponent consistent with the classical percolation study ($\nu = 4/3$) of two-dimensional disordered systems \cite{PhysRevLett.58.2325, PhysRevB.27.7539, RevModPhys.67.357}. We argue that such percolation limit applies, given the moderate disorder strengths corresponding to QPTs, with random potentials varying slowly over the system size, thus reducing quantum tunneling and interference effects \cite{10.1063/1.5086408}. Namely, we speculate that the clustering of trivialized real-space puddles, as described by classical percolation, is responsible for the value of $\nu$. 

We stress that the scaling is universal for the phases with stable 
Euler semimetallic nodes independent of the number of bands, as long as the band dispersion does not introduce band pockets overlapping with the nodal energies within the disorder width.
Namely, we corroborate these results of similar gap-closing QPTs in three-band models with ${\chi = 2}$ \cite{_nal_2020, multigap}, hosting four stable nodes between the bottom bands, as well as in four-band Euler models. 
For the latter this concerns Hamiltonians $H^{(\chi_1,\chi_2)}(\boldsymbol{k})$ having two Euler-characterized nodal two-band subspaces separated by a gap~\cite{multigap},  rendering similarly stable Euler semimetals with invariants denoted by $\chi_{1,2}$.

\begin{figure}[t]
\includegraphics[width = \columnwidth]{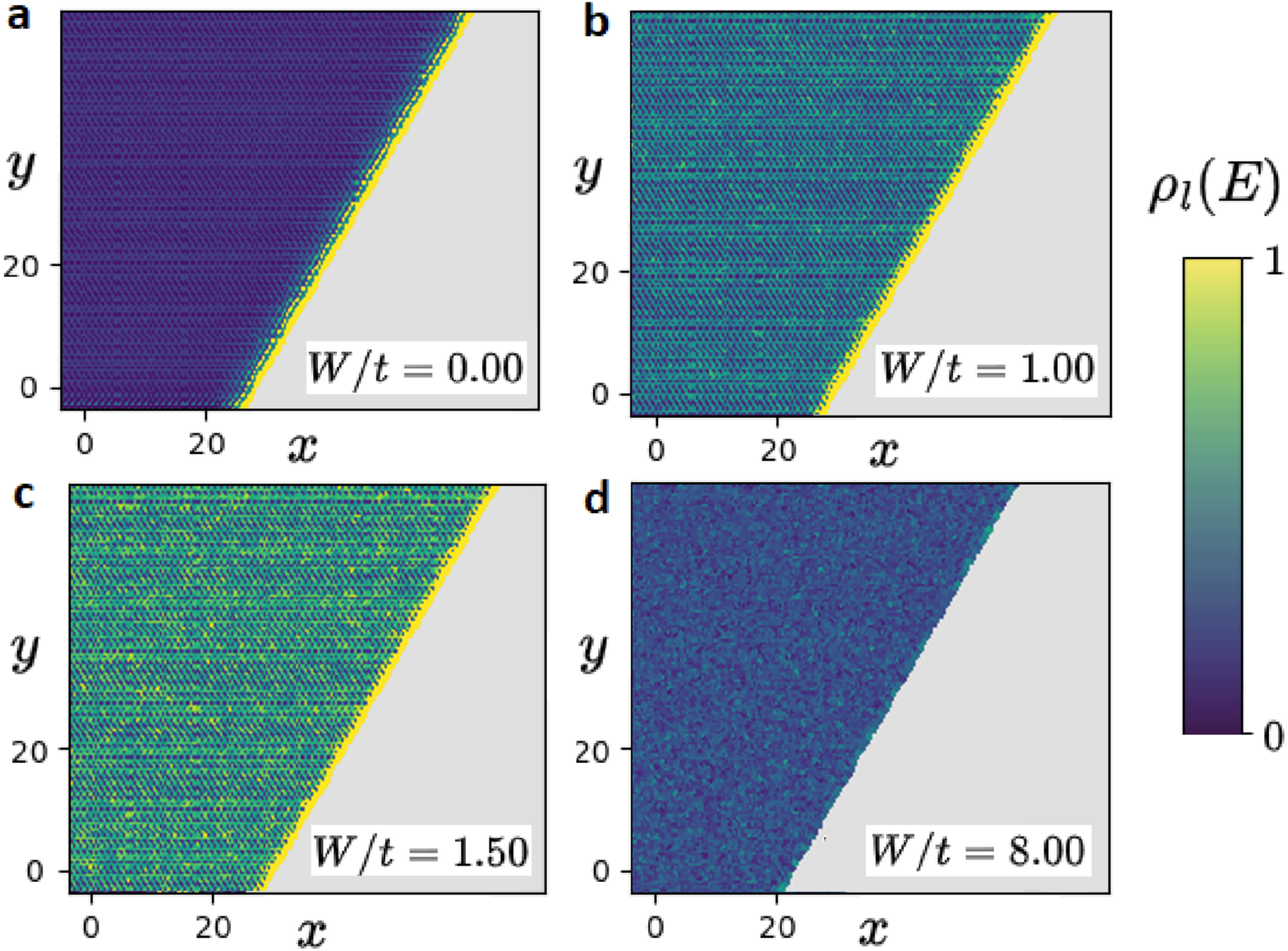}
\caption{Disorder-averaged LDOS around the edge (transition to grey areas) of kagome Euler semimetal with $\chi = 1$, at the clean-phase energy of protected nodes, for 200 disorder realizations. The evolution of edge states with disorder is plotted for the model kagome phase $(t,t') = (1,-0.2)$, with mass term opening the upper gap providing a protection from unbraiding. Contrary to the finding in Weyl semimetals \cite{PhysRevB.96.201401}, where dissolution of edge states happens already at subcritical disorders, we observe the stability of edge states from undisordered Euler semimetal \textbf{(a)} up to the upper-gap-closing critical disorder ($W_c = 1.05\pm0.05$), via an unbraiding regime \textbf{(b)}, prior to a QPT to a metal \textbf{(c)}, which at much higher disorder is followed up by Anderson localization \textbf{(d)}. We observe that the edge state due to the $\pi$-Zak phase persists throughout the entire unbraiding regime, before the bulk states become delocalized on percolation clustering (see also App.~\ref{app:appC}).}
\label{fig:fig3}
\end{figure}

\sect{Edge states protected by braiding}
To complete the study, we also investigate the effects on edge states accompanying the bulk. Similarly to Chern insulators \cite{PhysRevLett.61.2015} with chiral edge modes, Euler semimetal Hamiltonians can support anomalous conductivity (see Fig.~\ref{fig:Kagome_xy}, App.~\ref{app:appD}). However, the edge states are present in multiple gaps \cite{RJ_Kagome}, at the energies that do not necessarily cross the Fermi level ($\mu \equiv E_F$, as we strictly consider systems at zero temperature, $T = 0$) and generically arise, as mentioned, by different mechanisms, such as a non-trivial Zak phase~\cite{RJ_meron,RJ_Kagome,Peng2021}. The dependence of the average $\sigma_{xy}$ on doping and disorder is shown in App.~\ref{app:appD}, Fig.~\ref{fig:Kagome_xy}. Contrary to the Chern insulators, the edge modes in Euler phases are not quantized by the associated topological invariant \cite{PhysRevB.25.2185}, but, in degenerate flatband limit, they can be thought of as a pair of helical edge states, where each branch is associated with a bulk mirror Chern number \cite{multigap,guan2022landau}. To track their evolution, we perform a local density of states (LDOS) calculation, indicating the fractional contributions of atomic orbitals $\ket{i}$ to the eigenstates at selected energies
\\
\begin{equation}\label{dospersite}
    \rho_{\textit{l}}(E) = \sum_{n} \abs{\bra{\kappa, \boldsymbol{i}}\ket{n}}^2 \delta(E - E_n),
\end{equation}
where $\kappa$ denotes an orbital at the site $\boldsymbol{i}$ of interest, and $\ket{n}$ represents an eigenstate of the Hamiltonian with energy $E_n$.

In Fig.~\ref{fig:fig3}, the LDOS plots corresponding to increasing disorder in the introduced stable Euler semimetal model were shown. Importantly, we find the braiding-assisted stability of edge states up to the full criticality, contrary to the finding in Weyl semimetals \cite{PhysRevB.96.201401}. 

\begin{figure}[t]
\includegraphics[width=\columnwidth]{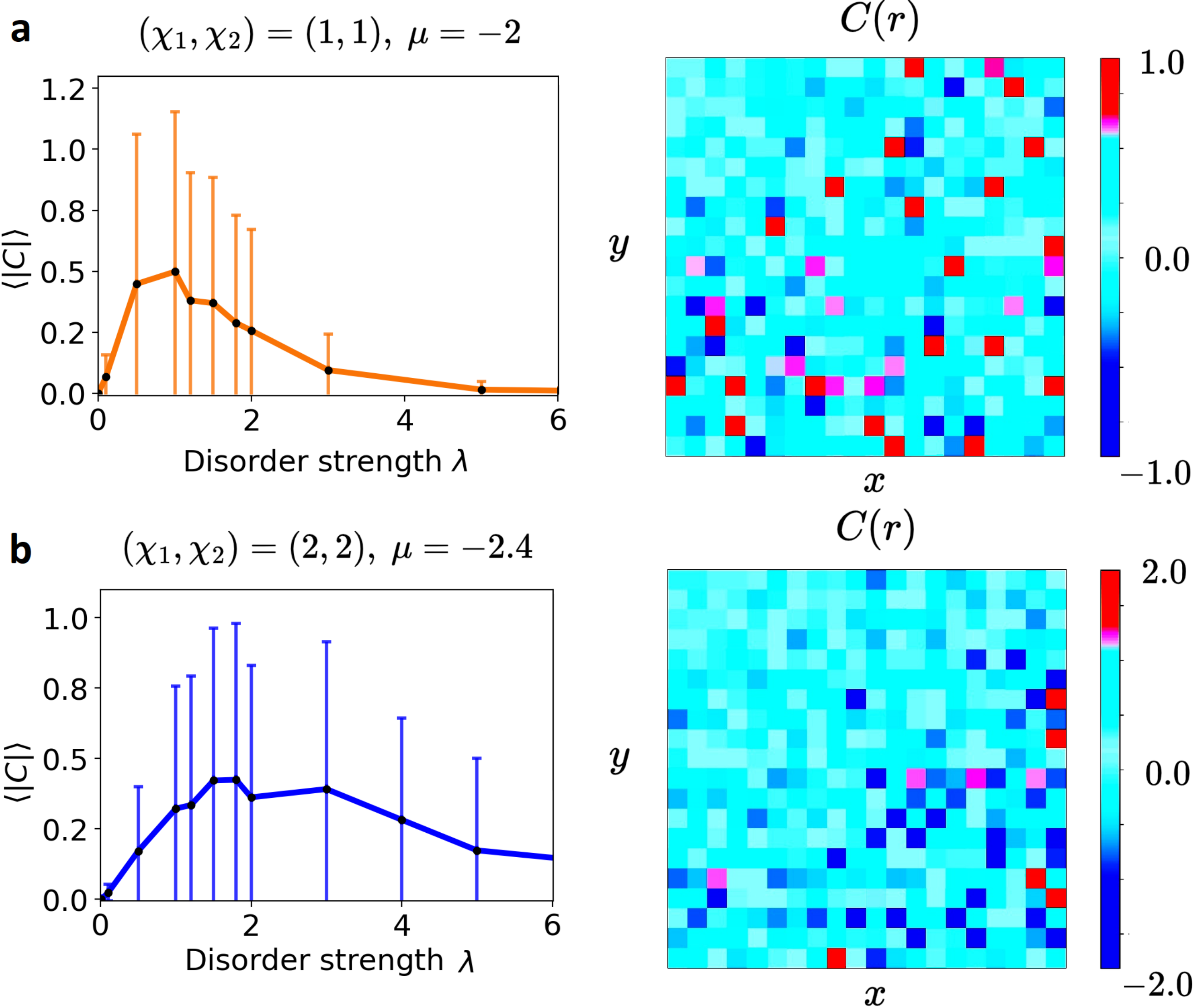}
\caption{Evolution of the averaged local Chern numbers (LCN) and their magnitudes $\langle\abs{C}\rangle$ with the strength of magnetic disorder [${\lambda = \text{max}(\lambda_i)}$] for different four-band Euler phases: \textbf{(a)}~${(\chi_1, \chi_2) = (1,1)}$, \textbf{(b)} $(\chi_1, \chi_2) = (2,2)$. The error bars correspond to the standard deviation in LCN. The landscape of Chern plaquettes with non-vanishing locally averaged Chern markers $C(\textbf{r})$ surrounded by trivial phase regions with $C(\textbf{r}) \approx 0$. The markers were evaluated over $100 \cross 100$ unit cells of the systems partitioned into $5 \cross 5$ puddles, at disorders corresponding to the peaks of $\langle\abs{C}\rangle$. We note that the peak in the LCN average of the $(\chi_1, \chi_2) = (2,2)$ phase is suppressed due to the presence of an extensive trivial region (\textit{cyan}).}
\label{fig:fig4}
\end{figure}

\sect{Islands with Chern topology}
In addition to the analysis of electronic properties and QPTs, we may similarly investigate the topological character of disordered Euler phases with Chern markers \cite{Kitaev20062, PhysRevB.84.241106, Caio2019} (see App.~\ref{app:appC}). The Chern marker study is motivated by the Bloch bundle complexification relations between Euler and Chern characteristic classes~\cite{RJ_ZrTe,multigap}, which can be further explored with disorder. We show how the markers indicating Chern topology locally in real space change on increasing disorder strength in different phases, see Fig.~\ref{fig:fig4}. As markers evaluate the imaginary part of the trace obtained from the real operators and eigenstates (see App.~\ref{app:appC}), enforced by real hoppings and onsite energies, these need to vanish in all Euler semimetals unless $\mathcal{T}$ symmetry is explicitly broken. \black{However, as we show in this section, a~magnetic disorder that breaks time-reversal and $\mathcal{C}_2\mathcal{T}$ symmetries in an Euler semimetal can in principle yield puddles of Chern phases. We find that the disorder inducing such Chern insulator islands is not arbitrary, and should take an effective form of a mass term effectively gapping out the nodes, i.e.~providing a gap in the DOS for given disorder realization. Before addressing this problem further, we elaborate on the use of a Chern marker to reflect the presence of Chernful islands surrounded by trivial regions in magnetically-disordered Euler semimetals. First, we importantly note that the Chern marker is directly related to the bulk anomalous Hall conductivity (AHC)~\cite{PhysRevB.95.121114}, which as an anomalous transport property for an insulating state supporting gapped DOS, is a direct probe for a plaquette of a Chern insulator. Here, the emergence of the Chern islands with quantized AHC can be intuitively understood as a phase separation of an Euler semimetal into trivial insulator and Chern insulator regions. Namely, the magnetic disorder considered above favours the development of local orbital magnetization aligned along the effective magnetic field imposed by the correlated disorder within an island. We note that a similar phase separation mechanism can be intuitively supported even by classical magnetic systems; applying an effective magnetic field by imposing a magnetic disorder in a paramagnet can locally orient the magnetic moments along the direction of the imposed magnetic field. In addition to this observation, we stress that the local orbital magnetization, which constitutes the entirety of the magnetization here, given that the models considered in this work are spinless (and therefore there is no spin magnetization), is in one-to-one correspondence with the local AHC, as shown in Ref.~\cite{PhysRevLett.110.087202}. Crucially, in an insulator, with no Fermi surface (as reflected by the gapped DOS at the Fermi level), AHC can only be associated with a quantized Chern number, which concludes the explanation of the mechanism and of the importance of the Chern markers demonstrated in Fig.~\ref{fig:fig4}. Here, specifically, the generation of Chern plaquettes on explicit time-reversal symmetry breaking induced by disorder} is achieved by adding real-space dependent local perturbations, e.g. of the form $\Gamma_{20}=\sigma_{2}\otimes\sigma_0$, to the four-band Bloch Euler model $H^{(\chi_1,\chi_2)}(\boldsymbol{k})$ (see App.~\ref{app:appA}), that is 
\\
\begin{equation}
\begin{aligned}
    H[\{\lambda_{\boldsymbol{i}}\}] &= \sum\limits_{\boldsymbol{i},\boldsymbol{j}} \sum\limits_{\alpha,\beta} t_{\alpha\beta}(\boldsymbol{i}- \boldsymbol{j}) c_{\boldsymbol{i},\alpha}^{\dagger} c_{\boldsymbol{j},\beta}\,,\\
    t_{\alpha\beta}(\boldsymbol{i}- \boldsymbol{j}) &= \sum\limits_{\boldsymbol{k}} e^{i \boldsymbol{k}\cdot (\boldsymbol{i}-\boldsymbol{j})}
    [H^{(\chi_1,\chi_2)}(\boldsymbol{k})]_{\alpha\beta} + \la_{\boldsymbol{i}} \delta_{\boldsymbol{i},\boldsymbol{j}} \,[\Gamma_{20}]_{\alpha\beta}\,.
\end{aligned}
\end{equation}
\\
Here $H^{(\chi_1,\chi_2)}(\boldsymbol{k})$ is free of disorder, $\alpha$, $\beta$ are orbital indices at sites labeled with lattice vectors $\boldsymbol{i}$, $\boldsymbol{j}$, and 
$\la_{\boldsymbol{i}} \in \small[-W, W \small]$ denotes the $local$ magnetic disorder strength that varies randomly from a site to another.

In such a case, the magnetic impurities  controlled by $\la_{\boldsymbol{i}}$ 
induce the opening of an energy gap in each two-band Euler subspace with $\chi_n$---i.e. hosting $2\chi_n$ stable nodes in the limit of clean crystal \cite{PhysRevB.102.115135}---leading to two gapped Chern subspaces with $C^{\pm}_n=\pm\chi_n$, as was studied in the case of clean models in Ref.~\cite{multigap}. Finally, we note that AHC also arises in Euler phases subject to the more arbitrary $\mathcal{T}$-breaking disorders, as we verify with KPM. However, AHC is not quantized under gapless conditions where Chern topology is not well-defined.

\sect{Discussion}
Compared with graphene, or Weyl semimetals of higher dimensionality, the results indicate a distinct quantitative critical behavior of non-trivial Euler phases upon adding disorder. The scaling of DOS, as well as the LDOS calculations in kagome phases show that Euler semimetals become naturally metallic at criticality, similarly to the findings in Weyl semimetals \cite{PhysRevB.96.201401}.
The scaling analysis suggests that the effects of disorder on two-dimensional Euler phases do not depend significantly on the finite-size, above the system sizes studied in this work.

As we show in four-band Euler models, the critical phases with local Chern numbers (LCNs) in real space can be obtained via QPTs induced by disorder, as long as both $\mathcal{C}_2\mathcal{T}$ and $\mathcal{T}$ symmetries are \textit{locally} broken in real space puddles in the system, e.g., with magnetic doping. That is, our findings show that islands with LCNs due to single-band topology can result from multi-band invariants on removing associated nodal structures with symmetry-breaking disorder. We note that if $\mathcal{T}$ symmetry is broken by construction in the clean $\mathcal{C}_2\mathcal{T}$-symmetric Hamiltonian of choice, even upon adding a non-magnetic quenched disorder respecting $\mathcal{C}_2\mathcal{T}$ symmetry on average, LCNs can be induced, as Ref.~\cite{wang2023anderson} suggests. This result can be heuristically understood as lifting the $C_2\mathcal{T}$-imposed degeneracy of two initial $\cal{T}$-breaking Chernful bands. In general, these kinds of transitions can be ascribed to the fragile topology, such as captured by the Euler class, as long as $\mathcal{T}$ is broken, while the parent invariant-protecting $\mathcal{C}_2\mathcal{T}$ is being preserved on average~\cite{PhysRevX.13.031016}. We reiterate that the notion of a disorder-induced unbraiding transition trivializing the Euler topology of semimetals cannot be captured by conventional symmetry indicators and that fragile topology in this context means that a gap closing with a trivial band can trivialize the topology (contrary to the strong topology necessitating a gap closing with a band with opposite invariant, e.g.~Chern number)~\cite{PhysRevB.102.115135}. For example, the mass term opening the upper gap can explicitly break $C_6$-symmetry of the kagome lattice, while preserving the Euler class, hence leaving the multi-gap topology captured by this characteristic class of real vector bundles unaffected.

\sect{Summary and outlook}
We demonstrate that Euler semimetals generically show a larger robustness to disorder, manifested by the enhanced stability of edge modes, than semimetals with trivial nodes. We corroborate these findings by studying the ADOS, LDOS, and conductance scaling with increasing disorder. We find that disorder-induced QPTs of Euler semimetals to metallic phases are universally characterized by the same localization length exponent, provided no pockets hinder probing the nodal structure. We argue that the underlying mechanism of QPT can be captured by multi-gap unbraiding mechanism on closing a neighboring gap, explaining the extended critical regime and stability of edge states. Finally, using topological markers, we also show that non-trivial Chern numbers can emerge in disordered Euler phases, upon breaking Hamiltonian symmetries locally, which could possibly originate from the presence of magnetic impurities in the system. While realizing Euler semimetals in electronic materials is an ongoing pursuit spurred on by recent progress in terms of metamaterial realizations~\cite{RJ_Kagome}, cold atoms and trapped ion insulators~\cite{zhao2022observation}, as well as a wealth of predictions~\cite{PhysRevB.102.115135, RJ_meron,RJ_Kagome,RJ_ZrTe, PhysRevB.100.195135, PhysRevB.105.214108, Peng2021, PhysRevB.105.L081117,bouhon2023quantum, PhysRevX.9.021013,PhysRevLett.126.246801,ahnprl,Jiang1Dexp, _nal_2020,Guo1Dexp}, our study provides evidence for topological protection controlled by experimentally realizable braiding. We identify novel effects of disorder in this context, underpinning an important aspect of this new direction.
\\
\begin{acknowledgements}
    W.~J.~J. acknowledges funding from the Rod Smallwood Studentship at Trinity College, Cambridge. M.~N. acknowledges funding from the Trinity Henry-Barlow Scholarship and M.~C.~Payne. A.~B. has been partly funded by a Marie Sklodowska-Curie fellowship, grant no. 101025315, \black{and acknowledges financial support from the Swedish Research Council (Vetenskapsradet) (2021-04681).} R.-J.~S. acknowledges funding from a New Investigator Award, EPSRC grant EP/W00187X/1, as well as Trinity College, Cambridge. \black{W.~J.~J. thanks Jan Behrends for helpful discussions.}
    The tight-binding calculations and plots were obtained using the Kwant code~\cite{Groth_2014}. \\ 
    
    \textit{Note added:} In the process of writing we became aware of the work \cite{wang2023anderson} studying critical metallic phase in disordered $\mathcal{C}_2\mathcal{T}$-symmetric insulators with fragile topology captured by the non-trivial second Stiefel-Whitney class. Our results, obtained for different systems and from a different perspective, namely semimetals with Euler topology protected by the same symmetry, are consistent.
\end{acknowledgements}

\bibliographystyle{apsrev4-1} 
\bibliography{references.bib}

\appendix

\section{Euler Hamiltonians}\label{app:appA}

\black{As we elaborate on in detail in the main text}, in this work, we study the effects of disorder in two-dimensional $(d = 2)$ three-band and four-band Hamiltonians of non-trivial Euler class discussed in the introduction. The randomly-generated quenched disorder is implemented in real space by tuning individual onsite energies after changing the basis of the model Hamiltonians via Fourier transforms
\\
\begin{equation}
t_{\alpha\beta}(\boldsymbol{i}- \boldsymbol{j}) = \sum_{\textbf{k} \in \text{BZ}} e^{i(k_x x + k_y y)} H_{\alpha \beta}(\textbf{k}),
\end{equation}
\\
where the vector $ \boldsymbol{r} = (x,y) = \boldsymbol{i} - \boldsymbol{j}$ determines the displacement corresponding to the hopping between unit cells $\boldsymbol{i}$ and $\boldsymbol{j}$, with $\alpha,\beta = 1, 2, 3, (4)$ or equivalently $A, B, C, (D)$ denoting the basis orbitals. The onsite energies are given by $\varepsilon_i = t_{\alpha\alpha}(\boldsymbol{0})$ contrary to the intra-cell/intra-site hoppings $t_{\alpha\beta}(\boldsymbol{0})$, where $\alpha\neq \beta$. 

\subsection{Three-band models}
\begin{figure}[t]
\includegraphics[width=\columnwidth]{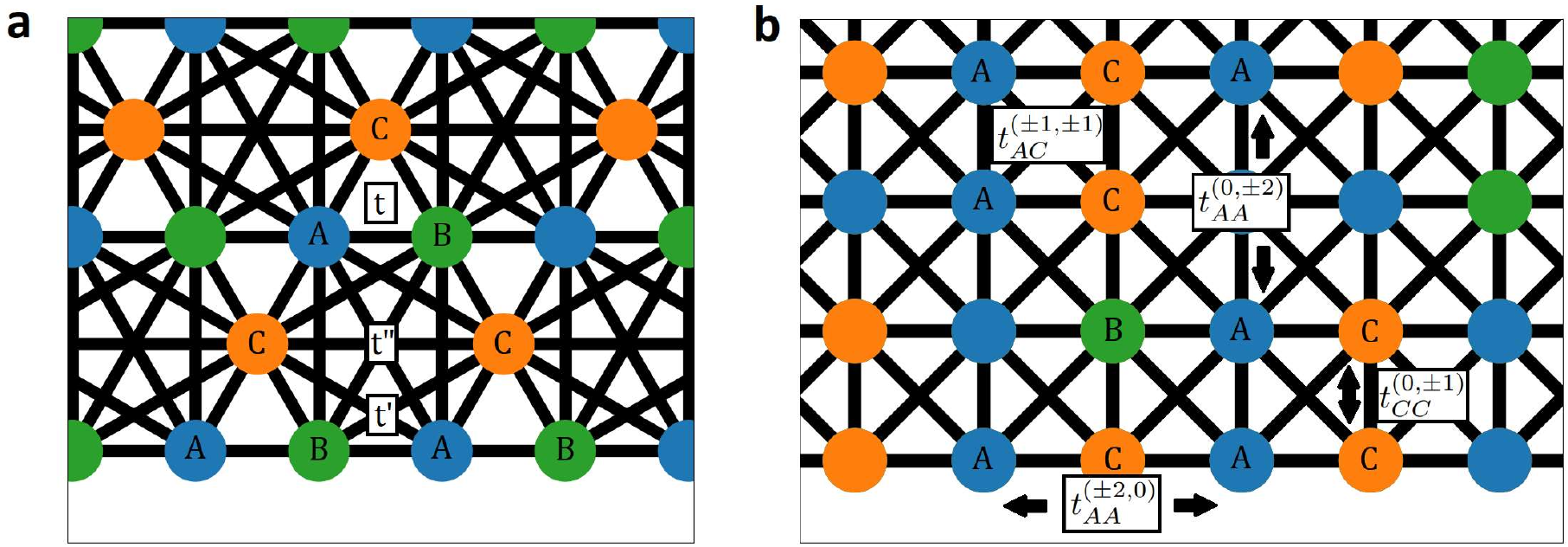}
\caption{\text{(a)} Kagome lattice adaptation of two-dimensional three-band Euler semimetals/insulators. \text{(b)} Square lattice realization of Euler Hamiltonians, as adapted for the four-band models. Colours represent distinct orbitals associated with different lattice sites.  Particular hoppings were marked with bold lines.}
\label{fig:lat}
\end{figure}

 The three-band models of Euler semimetals and insulators, most recently studied experimentally, consist of a kagome lattice \cite{RJ_Kagome, RJ_meron} hosting topological band nodes of $\chi = 1$ in the two-band subspace. The kagome lattice, see Fig.~\ref{fig:lat}, consists of three distinct lattice sites at Wyckoff positions $3c$ of hexagonal layer group L80, hosting orbitals with tunable onsite energies~\cite{RJ_Kagome}. The topological Euler phases can be obtained by adding properly parametrized nearest-neighbor $t$, next-nearest-neighbor $t'$, and importantly third-neigbour $t''$ (N3) hopping amplitudes to the Hamiltonian, respecting rotational $C_{6}$ symmetry of the lattice. The corresponding Euler Hamiltonians can be written as
\begin{equation}
\textit{H}(\textbf{k}) = 
\begin{pmatrix}
        H_{AA}(\textbf{k}) & H_{AB}(\textbf{k}) & H_{AC}(\textbf{k})\\
        H_{AB}(\textbf{k}) & H_{BB}(\textbf{k}) & H_{BC}(\textbf{k})\\
        H_{AC}(\textbf{k}) & H_{BC}(\textbf{k}) & H_{CC}(\textbf{k}) \\
\end{pmatrix},
\\
\end{equation}
\begin{align}
& \hspace{-1mm} H_{AA}(\textbf{k}) = \varepsilon_{A} - 2t'' \cosine{(k_1)}, \\
& \hspace{-1mm} H_{AB}(\textbf{k}) = -2t \cosine{(k_1/2+k_2/2)} - 2t' \cosine{(k_1/2-k_2/2)}, \\
& \hspace{-1mm} H_{AC}(\textbf{k}) = -2t \cosine{(k_2/2)} - 2t' \cosine{(k_1+k_2/2)}, \\
& \hspace{-1mm} H_{BB}(\textbf{k}) = \varepsilon_{B} - 2t'' \cosine{(k_2)}, \\
& \hspace{-1mm} H_{BC}(\textbf{k}) = -2t \cosine{(k_1/2)} - 2t' \cosine{(k_1/2+k_2)}, \\
& \hspace{-1mm} H_{CC}(\textbf{k}) = \varepsilon_{C} - 2t'' \cosine{(k_1+k_2)},
\end{align}

\black{with $k_1, k_2$ denoting the $\textbf{k}$-vector components along the reciprocal lattice vectors $\textbf{b}_1$, $\textbf{b}_2$, i.e. $\textbf{k} = \frac{k_1}{2\pi} \textbf{b}_1 + \frac{k_2}{2\pi}  \textbf{b}_2$}. As mentioned in the main text, the two-band Euler subspace can be isolated from the neighboring band by the addition of a diagonal mass term $\text{diag}(-m, -m, 0)$ respecting $C_2\mathcal{T}$, but breaking $C_6$ symmetry, e.g., on setting $m = 3$. Unlike in Euler insulators, the band structure of Euler semimetals hosts band nodes at Fermi level $E_F$, which is attainable by setting $t'' = 0$ \cite{RJ_Kagome}. Some of these nodes (e.g. with frame charges $\pm i$, see the main text) are analogous to the linear Dirac cones in graphene, which is a zero-gap semiconductor (i.e. a semimetal) as well. Three-band models with higher Euler class, such as $\chi = 2$ \cite{_nal_2020}, are extensively discussed in Ref.~\cite{multigap}

\subsection{Four-band models}

Beyond the three-band Hamiltonians we investigate disorder, including magnetic terms, in minimal four-band models with double Euler class associated with the presence of two topologically non-trivial two-band subspaces \cite{PhysRevB.102.115135}. The corresponding Hamiltonians can be generated following a general procedure, with Euler invariants induced in two-band subspaces via Pl\"ucker embedding \cite{PhysRevB.102.115135}. However, minimal models preserving the Euler topology can be constructed, while setting appropriate cutoffs on the number of hoppings. For a double Euler invariant, in the Bloch basis, these can be effectively expressed as \cite{multigap}
\\
\begin{equation}
\begin{split}
    H^{(1,1)}(\textbf{k}) = \sine{k_1}\Gamma_{01} + \sine{k_2}\Gamma_{03}
    - [m - t_1(\cosine{k_1} + \cosine{k_2})]\Gamma_{22} \\+ \delta\Gamma_{13},
\end{split}
\end{equation}
\begin{equation}
\begin{split}
    H^{(2,2)}(\textbf{k}) = \sine{k_1}\Gamma_{01} + \sine{k_2}\Gamma_{03}
   - [m - t_1(\cosine{k_1} + \cosine{k_2}) \\- t_2\cosine{(k_1+k_2)}]\Gamma_{22} + \delta\Gamma_{13},
\end{split}
\end{equation}
\\
where $\Gamma_{ij} = \sigma_{i} \otimes \sigma_{j}$ are $4 \cross 4$ Dirac matrices and $(m, t_1, t_2, \delta)$ is a set of tunable parameters. As introduced in Ref.~\cite{multigap}, representative models for $(\chi_1, \chi_2) = (1, 1)$ and $(\chi_1, \chi_2) = (2, 2)$ can be generated by $(1,-3/2, 0, 1/2)$ and $(1/2,-1/2, -3/2, 1/2)$ parametrizations, respectively.

\section{Euler invariant and unbraiding under disorder}\label{app:appB}

In this section, we elucidate how the Euler invariant can be defined under disorder, when $\textbf{k}$ is no longer a good quantum number associated with the single particle eigenstates, and therefore the definition introduced in the main text cannot be directly used. Furthermore, we provide a formulation and a quantitative discussion of the disorder-induced unbraiding introduced in the main text.

\subsection{Euler invariant in disordered systems}

Here, we elaborate on how the Euler topology under disorder can be defined. Similarly to the Chern topology, a natural strategy is to try defining indicators acting similarly to the Chern marker or Bott index in the Chernful systems. On performing analogous steps, to defining the Chern markers, as derived in the Ref.~\cite{PhysRevB.84.241106}, one ends up with
\\
\beq{}
    \chi(\textbf{r}) = \frac{2\pi}{A} \sum_{\textnormal{a} \in \textnormal{cell}} \bra{\textbf{r}_{a}}\hat{T}\hat{y}\hat{Q}\hat{x} - \hat{T}\hat{x}\hat{Q}\hat{y}\ket{\textbf{r}_{a}} = \frac{2\pi}{A} \textnormal{Tr}_{\textnormal{cell}} \{\hat{T}\hat{x}\hat{P}\hat{y}-\hat{T}\hat{y}\hat{P}\hat{x}\},
\eeq
for an Euler marker in a cell positioned at $\textbf{r}$ in terms of the orbitals $\ket{\textbf{r}_{a}}$, and projectors onto occupied $\hat{P} = \sum^{\text{occ}}_j \ket{E_j}\bra{E_j}$ and unoccupied $\hat{Q} = 1 - \hat{P}$ energy states, defined identically as for the Chern markers (see Sec.~III, \black{Numerical Methods}, App.~\ref{app:appC}). However, here distinctively, \black{an additional  ``transition"/ ``transfer"} operator $\hat{T}$ needs to be introduced in the formulation,
\\
\begin{equation}
    \hat{T} = \frac{A}{(2\pi)^2} \int_{\textnormal{BZ}} \dd^2 \textbf{k}~ \ket{\psi_{n+1,\textbf{k}} }\bra{\psi_{n,\textbf{k}}}.
\end{equation}
\\
Here, $A$ denotes the real-space unit cell area, while $n,n+1$ refer to the Euler band indices. We note that an issue for a general real-space formulation is that such transition operator would need to introduce a refined resolution between the Euler eigenstates of the disordered system. More specifically, as the disorder hybridizes multiple Bloch states at different $k$-points (see also the next subsection), the identities of the Bloch bands $\ket{\psi_{n,\textbf{k}} }, \ket{\psi_{n+1,\textbf{k}}}$ become lost.

\black{Nonetheless, these are known in the clean limit, threfore $\hat{T}$ can be heuristically defined before the disorder is added, and can be recast in the real space basis, which can be used as a reference also in the disordered system.} Namely, on Fourier transforming the Bloch states and expressing these in terms of the Wannier orbitals in the clean limit, we can express the operator as:
\\
\begin{equation}
    \hat{T} = \sum_{a,b} c_{ab} \ket{\textbf{r}_{a}} \bra{\textbf{r}_{b}},
\end{equation}
\\
where $c_{ab}$ encodes the products of the coefficients $c_{a,j}, c^{*}_{b,j'}$ obtained on replacing/identifying $\ket{\psi_{n,\textbf{k}} }, \ket{\psi_{n+1,\textbf{k}}}$ as eigenstates ${\ket{E_j} = \sum_{a} c_{a,j} \ket{\textbf{r}_{a}}}$ expressed in the real space orbital basis, which is naturally convenient for the disordered systems. Having formulated $\hat{T}$ in terms of the real space orbital coefficients, one can use $\chi(\textbf{r})$ as a proxy for the Euler topology in the disordered system, although similarly to the Chern markers, it is not guaranteed to yield quantized values. Nonetheless, by construction, such marker restores the quantization in the clean limit, where it coincides with the Euler invariant defined in the main text. In particular, we expect that averaging over the real-space cells within a total area $A_{\text{tot}}$ converges to the Euler invariant of a disordered system, \textcolor{black}{$\langle \chi(\textbf{r}) \rangle \equiv \frac{1}{A_{\text{tot}}} \int \dd^2\textbf{r}~\chi(\textbf{r}) \rightarrow \chi$. In Fig.~\ref{fig:marker}, we demonstrate the scaling of the marker with disorder in the Euler semimetal realized on the kagome lattice, as introduced in the main text. Nonetheless, we note that by making a reference to the clean limit, the marker defined here is a heuristic proxy for Euler topology in disordered system. To rigorously define the Euler invariant under disorder, we introduce a quantized definition under averaging over disorder realizations. Namely, we define the symmetry-averaged $\bar{\chi}$ Euler invariant from an ensemble/replica average of the disordered system which defines effective band structure parameters, which can be obtained within the Born approximation picture. In particular, we recognize that within the Born approximation, the onsite disorder effectively renormalizes the mass parameter entering the diagonal term. For uniform Anderson disorder of strength $W$, we find that by the action of disorder, the mass parameter $m$ is renormalized by self-energy $\bar{m} = m - \mathfrak{Re}~\Sigma_{11}$, where $\Sigma_{ij}$ denotes the elements of the self-energy matrix.
\begin{figure}[t]
\includegraphics[width = \columnwidth]{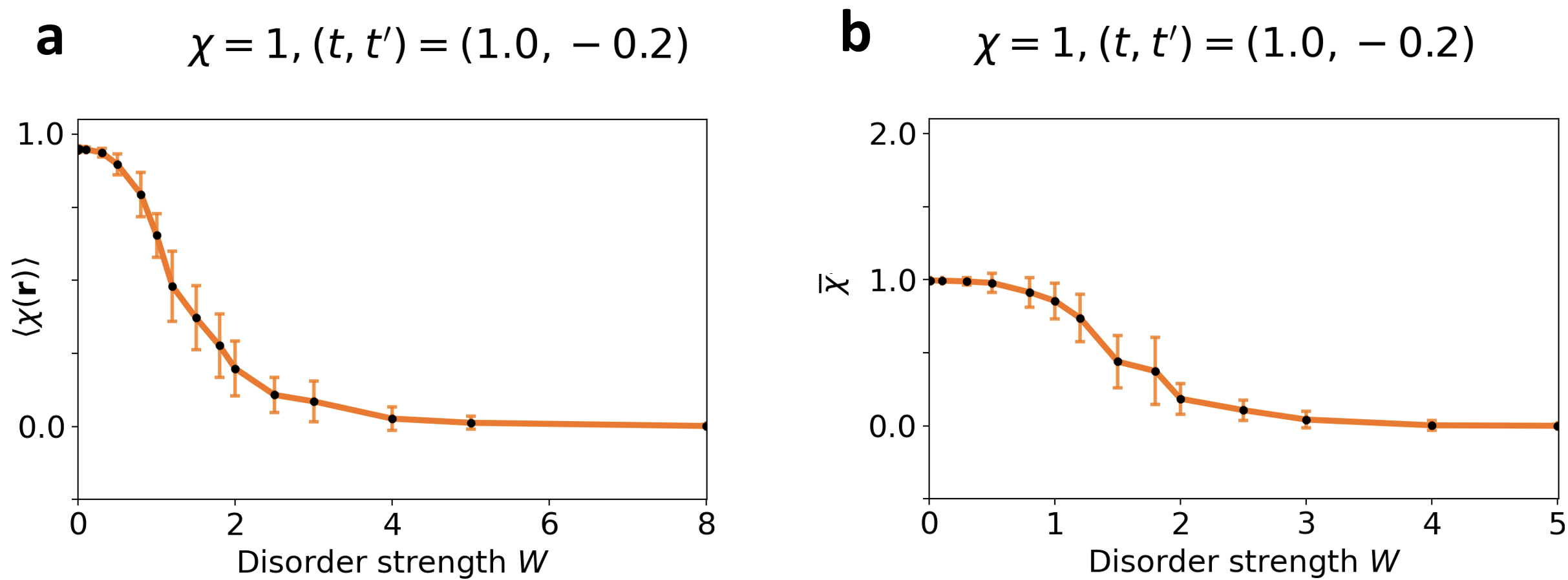}
\caption{\black{Pinpointing the topological phase transitions of the Euler invariant $\chi$ in the Euler semimetals under the uniform disorder of strength $W$.  (\textbf{a}) The scaling of an averaged Euler marker $\langle \chi(\textbf{r}) \rangle$ with its standard deviation plotted on top, against the disorder strength~$W$. Here, $\langle \chi(\textbf{r}) \rangle$ was averaged over the bulk realizing the kagome Euler Hamiltonian studied in the main text, with the system size $L \times L = 100 \times 100$ unit cells. (\textbf{b}) The average Euler invariant $\bar{\chi}$ defined under the disorder-averaging. Notably, the quantization of $\bar{\chi}$ is lost above the disorder value corresponding to the gap-closing in the effective band structure, as the Euler class $\chi$ [over the entire BZ] is quantized only if the gap above the Euler bands is preserved~\cite{RJ_ZrTe}. The error bars correspond to the numerical error in the computation of $\bar{\chi}$~\cite{RJ_ZrTe}, combined with the uncertainty in the effective mass on achieving the numerical self-consistency within the Born approximation. Both indicators of the Euler topology in disordered system approximately coincide, indicating the critical disorder ($W_c$) regime retrieved from the scaling analysis in the main text. Intuitively, the overlap of two real-space indicators can be understood as reflecting an expected coincidence of the ensemble average with a spatial average over a large single system realizing the Euler topology.}}
\label{fig:marker}
\end{figure}
The self-energy matrix is explicitly given by an integral self-consistency equation, which defines the self-consistent Born approximation (SCBA). Namely, for uniform disorder, we have,
\beq{}
    \Sigma = \frac{W^2 A}{3} \int_{\text{BZ}} \frac{\dd^2 \textbf{k}}{(2\pi)^2} \frac{1}{\mu+i0^{+} - H(\textbf{k})- \Sigma}.
\eeq
For the model Hamiltonian on a kagome lattice, as introduced in the main text, we set $\mu = -3.775$, further obtaining the self-energy $\Sigma$ matrix and correspondingly the effective mass term as a function of disorder $W$. On numerically satisfying the self-consistency condition, we find that effectively $\mathfrak{Re}~\Sigma \propto -W^2 \times \text{diag}(-1, -1, 0)$ in the weakly disordered limit. We present the dependence of an effective band structure on the effective mass term deduced from the self-energy in Fig.~\ref{fig:effbraiding}. Correspondingly, the average Euler class $\bar{\chi}$ can be evaluated from the two-band subspace of the effective bands, using the definition of $\chi$ from the main text, at any point of the evolution of the original Hamiltonian with the renormalized mass $\bar{m}$, which depends on the disorder strength $W$ (see Fig.~\ref{fig:marker}). In particular, at the critical disorder $W_c$, the average Euler class $\bar{\chi}$ suffers a discontinuous change in value from $\bar{\chi} = 1$ to $\bar{\chi} = 0$, as a result of the expicitly demonstrated unbraiding.}
\begin{figure*}[t]
\includegraphics[width = 0.9\linewidth]{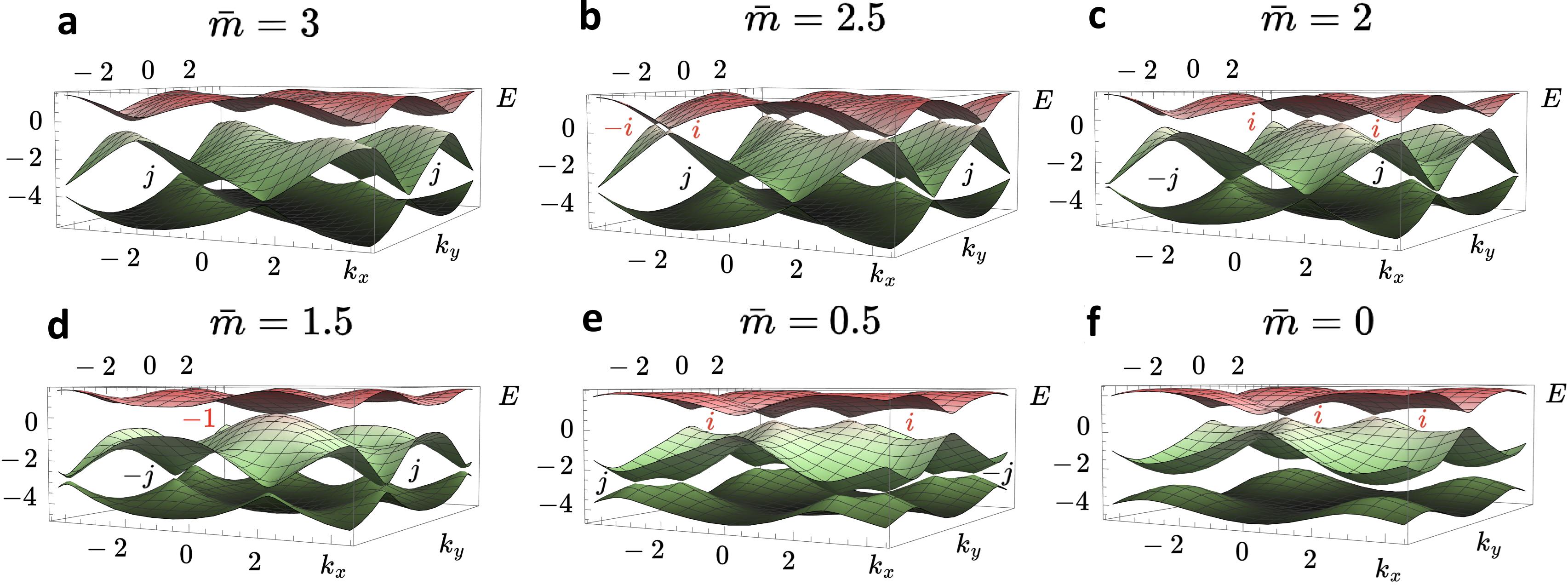}
\caption{\black{The evolution of the effective band structure with the effective mass $\bar{m}$ renormalized on the evolution of self-energy $\Sigma$, as a function of disorder strength $W$ under the disorder-averaging within SCBA. The renormalized mass induces the gap closure (\textbf{a}--\textbf{b}), and subsequent unbraiding (\textbf{c}--\textbf{d}), defined with effective band structure parameters obtained from SCBA. In this regime, as the effective mass $\bar{m}$ induced by the disorder changes, the nodes in effective band structure evolve in multi-gap braiding trajectories within two gaps, changing the relative signs of the nodal quaternion charges. Correspondingly, the unbraiding results in the trivialization of the disorder-averaged Euler invariant $\bar{\chi}$ that can be directly deduced from the corresponding effective eigenvectors. In the absence of the non-trivial Euler class, the nodes in the lower gap are no longer protected from annihilation (\textbf{e}) and can be gapped out at the higher effective masses, as shown in (\textbf{f}).}}
\label{fig:effbraiding}
\end{figure*}

\subsection{Unbraiding under disorder}

In the clean limit, the eigenstates in band $n$ are ${\ket{\psi_{n,\textbf{k}}} = e^{i\textbf{k} \cdot \textbf{r}} \ket{u_{n,\textbf{k}}}}$, as follows from the Bloch's theorem. Any given disorder realization can be decomposed in terms of the Fourier modes $V(\textbf{r}) = \sum_\textbf{q} V_\textbf{q} e^{i\textbf{q} \cdot \textbf{r}}$, which can be treated as a perturbation of the Hamiltonian in the clean limit $H$. With the disorder values inducing corresponding Euler semimetal-metal quantum phase transitions studied in the main text being weak, contrary to the strong disorder values required to cause the Anderson transitions, we assume that for capturing the unbraiding, we can treat disorder perturbatively to first order. Thus, as the perturbation is not time-dependent, but the band degeneracies are generically present, one can include disorder within the time-independent \textit{degenerate} perturbation theory, yielding a set of secular equations in a form of the determinant to solve for the energy eigenvalues ${E = E_1, E_2, \ldots, E_j, \ldots}$.
\begin{align}
\black{\Big|\Big|\sum_{\textbf{q}} \underbrace{\bra{u_{m,\textbf{k}'}}} _{i} V_\textbf{q} e^{i(\textbf{q}-\textbf{k}'+\textbf{k}) \cdot \textbf{r}}  +  e^{-i\textbf{k}'\cdot \textbf{r}} H e^{i\textbf{k}\cdot \textbf{r}} \underbrace{\ket{u_{n,\textbf{k}}}}_{j} - E \delta_{ij}\Big|\Big| = 0,}
\end{align}
with $\delta_{ij}$, the Kronecker delta for eigenstate labels $i,j$. We define $\ket{\tilde{\psi}_{n,\textbf{k}}} \equiv \ket{E_j}$, such that the overlap $\Big|\bra{E_j}\ket{\psi_{n,\textbf{k}}}\Big|$ found in terms of the orbital coefficients $c_a$, is maximized. In particular, if the overlap values are identical for multiple states, one should select a maximally-smooth reconstruction of $\ket{\tilde{\psi}_{n,\textbf{k}}}$ viewed as vector-valued functions $\ket{\tilde{\psi}_{n,\textbf{k}}} = (\underbrace{c_{a}(\textbf{k})}_{c_{a,j}}, \ldots,\underbrace{c_{N}(\textbf{k})}_{c_{N,j}})^T$ over $\textbf{k}$ in BZ. Here, $N$ is the total number of orbitals in the finite-size system.  

Correspondingly, the energies of the perturbed eigenstates were changed by disorder as $\tilde{E}_{n}(\textbf{k}) = E_j$. In the clean limit, $V_\textbf{q} \rightarrow 0$, the so-defined $\tilde{E}_{n}(\textbf{k})$ perturbative eigenstate energies naturally restore the original band structure.

Finally, the unbraiding can be deduced from tracking the perturbed energies $\tilde{E}_{n}(\textbf{k})$ of the perturbed states with indices $n$ and $\textbf{k}$, on representing these in a one-to-one correspondence to a band structure of the clean limit $E_{n}(\textbf{k})$. Such an operation, tracking the descendancy of the clean eigenstates on adding disorder, can be performed despite $\textbf{k}$ not being a good quantum number anymore.

It should be emphasized that while $n$ and $\textbf{k}$ are present as labels in perturbed eigenstates $\ket{\tilde{\psi}_{n,\textbf{k}}}$, these labels simply act as a state index tracking the evolution of a given eigenstate, whereas $\textbf{k}$ is no longer a good quantum number, as the disorder mode coefficients $V_\textbf{q}$ are switched on as perturbations. The overlap criterion allows to reconstruct the necessary correspondence, i.e. $\ket{\psi_{n,\textbf{k}}} \rightarrow \ket{\tilde{\psi}_{n,\textbf{k}}}$, from \black{maximally-overlapping}, real space basis coefficients $c_{a,j}$. Importantly, these coefficients do not obey the translational symmetry in~general, after the disorder is added.

\textcolor{black}{More transparently, beyond the paradigm of the energy spectrum of an individual disorder realization, the unbraiding follows from the disorder-ensemble averaged picture within the Born approximation. Namely, by deducing the effective mass parameter $\bar{m}$, from the self-energy $\Sigma$ as a function of the disorder strength ($W$), the effective band structure can be plotted (see Fig.~\ref{fig:effbraiding}). Here, the multi-gap unbraiding is explicitly demonstrated within the effective band structure. It should be noted that although every single disorder realization breaks the translational symmetry, the disorder-averaging within SCBA effectively restores the quasimomentum $\textbf{k}$ as a quantum number conserved under the scattering of quasiparticles from the disorder potential, along with the renormalization introduced with the self-energy. In particular, the momentum conservation is effectively restored as the impurity averaged electron propagator (Green’s function) becomes diagonal in momentum space after the impurity/disorder averaging~\cite{Coleman_2015}.}

\section{Numerical Methods}\label{app:appC}

\subsection{Kernel polynomial method (KPM)}

\begin{figure}[t]
\includegraphics[width = \columnwidth]{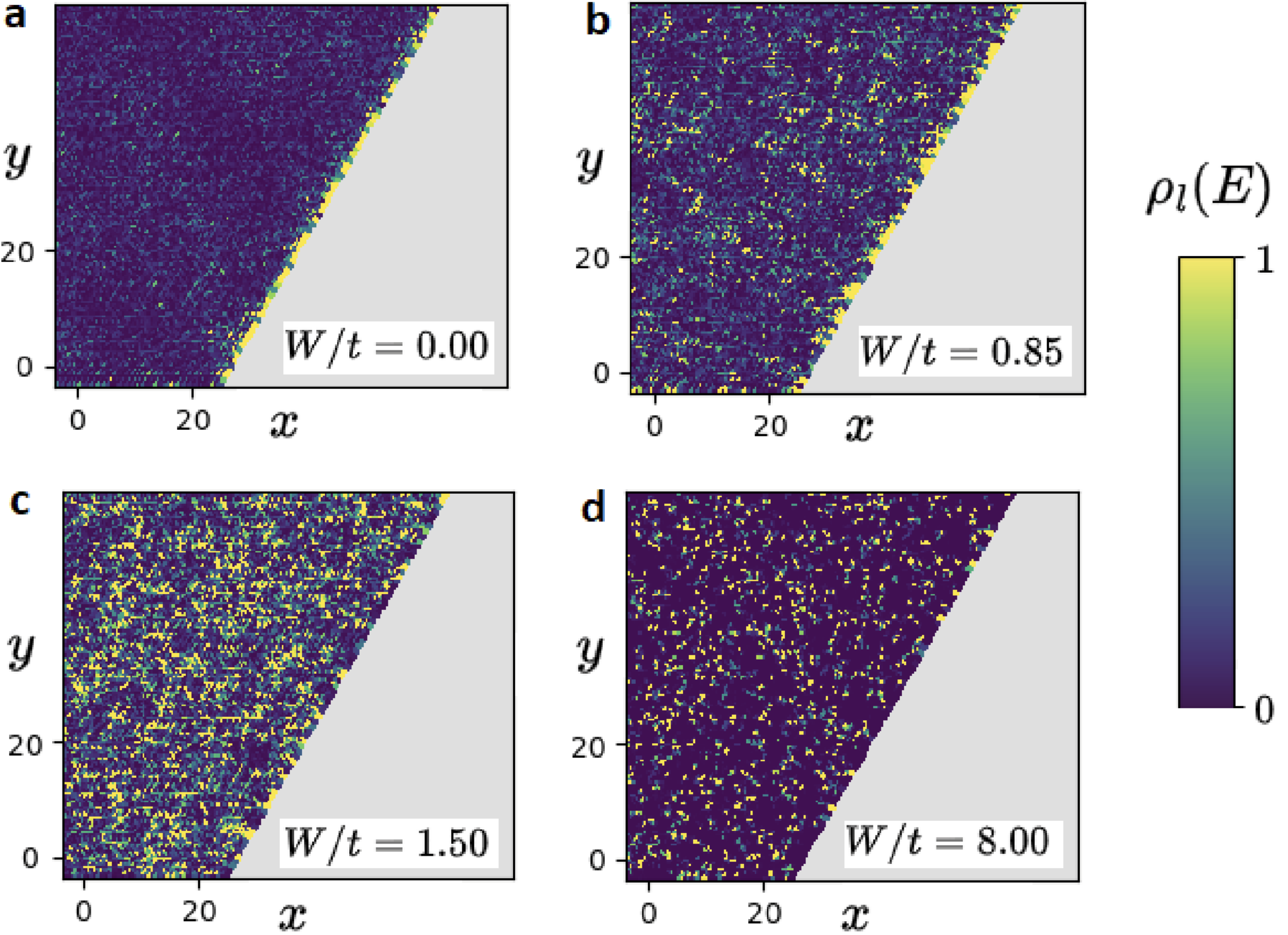}
\caption{LDOS around the edge (transition to grey areas) of kagome Euler semimetal with $\chi = 1$, at the clean-phase energy of protected nodes, for a single disorder realization. The evolution of edge states with disorder is plotted for the kagome phase with $(t,t') = (1,-0.2)$ and mass term opening the upper gap, which provides a protection from unbraiding. We observe the stability of edge states from clean Euler semimetal \textbf{(a)} up to the upper-gap-closing critical disorder ($W_c = 1.05\pm0.05$), contrary to the finding in Weyl semimetals \cite{PhysRevB.96.201401}, where dissolution of edge states happens already at subcritical disorders. Namely, we find that the edge state due to the $\pi$-Zak phase persists throughout the entire unbraiding regime \textbf{(b)}, prior to a QPT to a metal, where the bulk states become delocalized and metallic on percolation clustering, which is clearly visible in the individual disorder realizations \textbf{(c)}. At much higher disorder, this is followed up by Anderson localization \textbf{(d)}. We note that the profile of Anderson-localized wavefunction (as captured by the LDOS in a single disorder realization) directly reflects the transition to an Anderson insulator, which occurs at strong disorder. The findings can be compared with the disorder-averaged study of the edge state stability included in the main text.}
\label{fig:LDOSsingle}
\end{figure}

For each model, we evaluate the corresponding DOS of finite-sized Euler phases averaged over 200 disorder realizations using the KPM \cite{KPM2006} implemented in the Kwant code \cite{Groth_2014} with 2048 Chebyshev moments $\mu_n$ and a Jackson kernel \cite{JacksonK}. The uniform disorder is imposed by random potential $V(r)$ with sublattice averages given by onsite energies of the phase without any disorder $\varepsilon_i = \Big< V(r) \Big>$, provided by the uniform probability distribution in the interval $\left[\varepsilon_i-W, \varepsilon_i+W \right]$. Additionally, we study Gaussian disorder with correlations given by
\\
\begin{equation}
\Big< V(r) V(r') \Big> = \frac{W}{\xi^2_c} \text{exp}\Big(-\frac{|r-r'|^2}{2\xi^2_c}\Big),
\end{equation}
\\
where the standard deviation of the distribution is proportional to the correlation length of disorder $\xi_c$. We set $\xi_c = 4a$, where $a$ is a lattice constant of the studied model, in analogy to the previous work \cite{PhysRevX.8.031076}. We note that as long as $\xi_c << L$, where $L$ is the length-scale characterizing the system size, the correlation length has no impact on the qualitative effects of disorder. Additionally, we evaluate the conductivity tensors averaged over 200 disorder realizations using Kubo-Bastin response theory \cite{BASTIN19711811} implemented via the KPM \cite{Disordered2015}, which allows us to study longitudal conductivity (Fig.~\ref{fig:CondXX}) of the bulk as well as anomalous conductivity (Fig.~\ref{fig:Kagome_xy}), which can be related to edge states. Finally, we also perform LDOS calculations for Euler semimetals, as described in the main text, to study the dissolution of edge states. We perform averaging over the edge for 200 disorder realizations, on setting the systems size to $L \times L = 100 \times 100$ unit cells. The dissolution of edge states in the critical metallic phase can be contrasted with the subcritical dissolution of edge states in Weyl semimetals, i.e. Fermi arcs, as discussed and explicitly shown in the corresponding figures in Ref.~\cite{PhysRevB.96.201401}. Additionally, for further comparison, we include LDOS data obtained for a single disorder realization (Fig.~\ref{fig:LDOSsingle}).

\subsection{Scaling analysis}

In this section, we explain how the scaling analysis at critical points of the Euler semimetal models is performed. First, we introduce reduced disorder strength $\delta = (W-W_c)/W_c$ where the critical disorder $W_c$ is found with the KPM ADOS calculations. The first scaling exponent of interest relates the localization length $\xi = |\delta|^{-\nu}$ to the disorder strength, where, $\nu$ is strictly positive, while $\xi$ should diverge at the quantum critical point. The second, dynamical scaling exponent $z$ characterizes correlation energy $E_0 \sim \delta^{\nu z}$ \cite{PhysRevX.8.031076}, which vanishes at criticality correspondingly. To deduce both critical exponents, we express the number of electronic states at energy $E$
\\
\begin{equation}
    N(E, L) = (L/\xi)^d G(E/\delta^{\nu z}, L/\delta^{-\nu}),
\end{equation}
\\
where $G$ is a universal scaling function and $L$ denotes the size of the system, counting the number of unit cells in each dimension. The density of states $\rho(E) = L^{-d}dN/dE$ yields definitionally
\\
\begin{equation}
\rho(E, L) = \delta^{\nu(d-z)} F(E/\delta^{\nu z}, L/\delta^{-\nu}),
\end{equation}
\\
where $F$ is another universal, though explicitly unknown, scaling function. From these relations, we conclude that for the DOS at nodal points corresponding to the Fermi level of the semimetals $\rho(0) \sim \delta^{\nu(d-z)}$, while at small energies comparable with correlation energies, $\rho(E) \sim \abs{E}^{\frac{d}{z}-1}$, where both can be deduced from KPM spectral results for average DOS \cite{PhysRevB.96.201401}. Additionally, we check the values of the dynamical scaling exponent from $\sigma_{xx} (\mu)\sim \mu^\frac{1}{z}$ conductivity scaling as a function of chemical potential, analogously to the scaling found in the study of Weyl semimetals \cite{Sbierski_2014, PhysRevB.92.115145}.

\subsection{\black{Chern markers and local Chern numbers}}

A topological invariant characterizing Chern insulators \cite{PhysRevLett.61.2015, PhysRevLett.49.405}, namely the Chern number ($C$), is well-defined as an integral over momentum space for electronic systems with translational symmetry. However, as disorder removes periodicity of the system, the integral corresponding to the invariant is no longer well-defined. Therefore, for studying properties of disordered materials, real-space indicators, independent of the $k$-space formulation, are needed. Additionally, as disorder removes the homogeneity of the sample, a local character of such topological markers needs to be ensured. The local Chern markers satisfying these conditions can be defined \cite{PhysRevB.84.241106, Caio2019} and can be recast into a simple commutator form~\cite{Kitaev20062}
\\
\begin{equation}
    C(\textbf{r}) = -\frac{2\pi}{A} \mathfrak{Im}~ \textnormal{Tr}_{\textnormal{cell}} \{\big{[}\hat{P}\hat{x}\hat{P},\hat{P}\hat{y}\hat{P}\big{]}\},
\end{equation}
\\
where the trace is evaluated over all orbitals in given unit cell of area $A$, $\hat{P}$ is a projector onto occupied states, and $\hat{x}$, $\hat{y}$ are position operators. Computationally, we evaluate average Chern markers $\langle C \rangle$, as well as their magnitudes $\langle \abs{C} \rangle$ and standard deviations, by sampling multiple unit cells with random vectors, averaging over a few $(20-25)$ disorder realizations. We adapt an implementation of projectors with occupations encoded via the KPM density of states calculation, as introduced in the previous studies \cite{PhysRevResearch.2.013229}. The Chern markers allow us to deduce the presence of Chern topology or a lack thereof when either non-magnetic or time-reversal symmetry breaking magnetic disorder is added to the Euler phases. To obtain the local Chern numbers (LCN), the markers are averaged over real space patches/plaquettes consisting of $5 \cross 5$ unit cells. As mentioned in the main text, we stress that a spectral gap separating occupied and unoccupied state manifolds should be present in the sampled disorder realizations, in order to ensure that the Chern marker and Chern topology are not ill-defined in the studied disordered system. Even though the gap can be arbitrarily small, it should nonetheless be present, as can be verified with density of states (DOS) calculations.

\section{\black{Additional} numerical results}\label{app:appD}

We here present more generic numerical results, as mentioned in the main text. Additional KPM data for the Haldane model~\cite{PhysRevLett.61.2015} of a Chern insulator and standard graphene (which we denote as the limit of the Haldane model with vanishing second-neighbor hopping $t' = 0$) were included for a further comparison with the kagome Euler semimetals~\cite{RJ_Kagome}. We symbolically denote uniform and Gaussian disorders as $W \equiv V_0$ and $W \equiv \sigma$, respectively.

Crucially, in kagome models, such as the one with $(t, t') = (0, 1)$ and $(t, t') = (1, 0)$, and considering a filling up to the nodes of {\it opposite}, non-stable, quaternion charges ($q=\pm i$) between the upper two bands, we find $z = 0.9 \pm 0.1$ and $\nu = 1.0 \pm 0.1$, where the error is mainly due to the different fitting for multiple system sizes ($L = 75, 100, 150$) and hence should be attributed to finite-size effects (Fig.~\ref{fig:lines}).
Generically we find that the ADOS of these unstable Euler semimetals changes less rapidly around nodal energies than in graphene, i.e. a larger critical disorder is required for a QPT to occur (Fig.~\ref{fig:DOS}). We also notice that on adding stronger, i.e.~higher than critical, disorder, the conductivities in such Euler semimetals always decay slower than in graphene Figs.~\ref{fig:CondXX},~\ref{fig:ScalingCond}), but as this occurs for phases with nodes of opposite topological charge, which can be annihilated to induce a gap, the effect might be attributed to the different number of nodes and dispersion effects, rather than to the band topology itself. As the nodes in these phases are not protected by a patch Euler class or a gap, with the Euler band subspaces carrying $\chi = 1$ around the $\Gamma$ point being fully occupied, this result overlaps with the scaling found in graphene, where $z = 1$ \cite{PhysRevLett.118.036601}.

\subsection{Average density of states evaluated with KPM}

\begin{figure*}
\includegraphics[width=\linewidth]{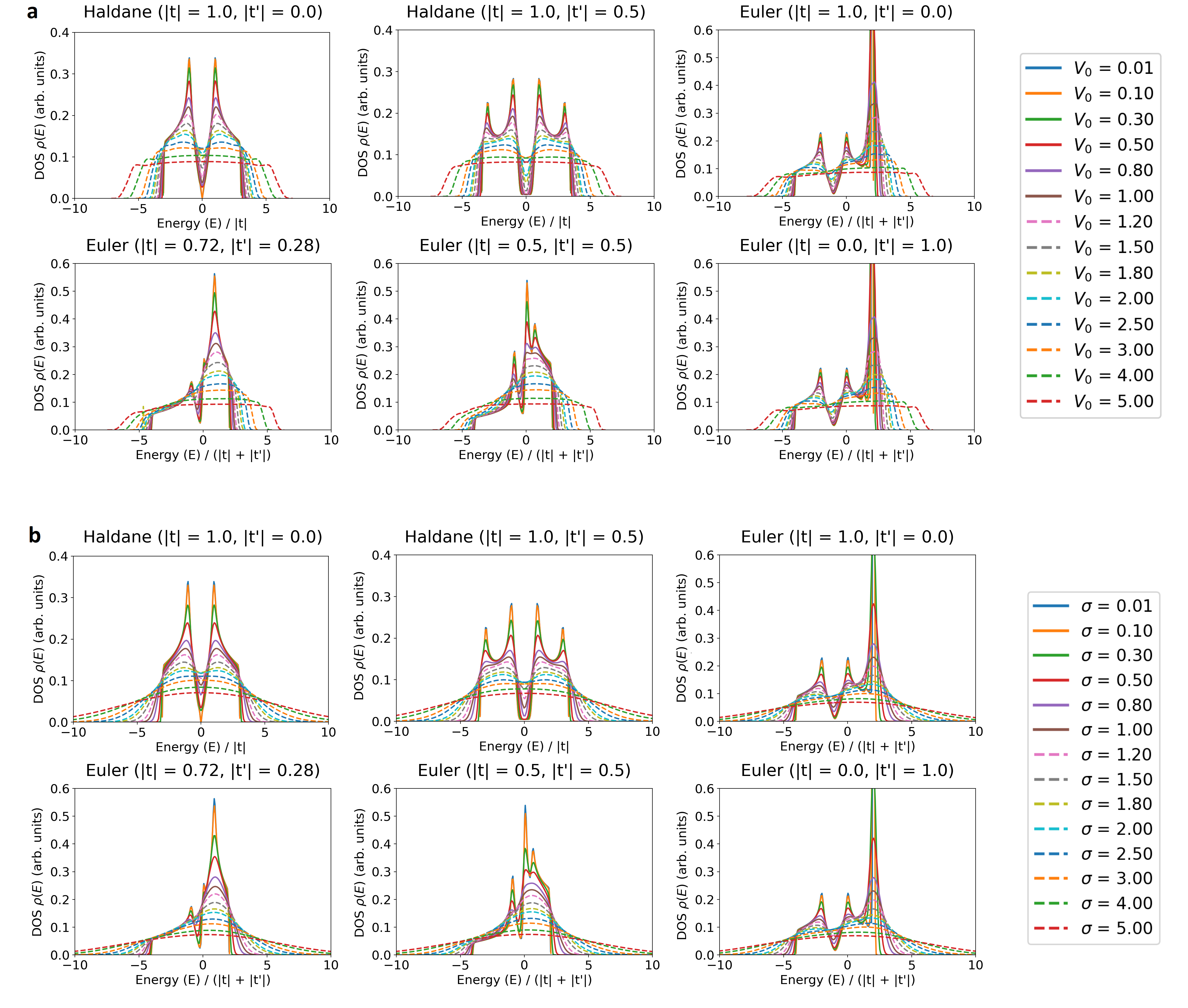}
\caption{ \textbf{(a)}: Average DOS of kagome Euler semimetals (and Haldane model, incl. its trivial phase --- graphene with nearest-neighbor hopping $t = 1$, next-nearest-neighbor hopping $t' = 0$, hence vanishing Chern number) subject to  \textbf{(a)}: uniform disorder, \textbf{(b)}: Gaussian disorder. As can be compared, we note that both distributions yield quantitatively almost identical DOS results, different only at high energies ($>5$) due to the boundedness of uniform disorder. At high disorder strengths $W$ ($> 5$), we observe strong localization with a flattened energy spectrum due to Anderson localized states, i.e.~the critical metallic phases obtained from Euler semimetals undergo quantum phase transitions to Anderson insulators. The plots were obtained for systems consisting of $L \cross L = 150 \cross 150$ unit cells.}
\label{fig:DOS}
\end{figure*}

We attach further ADOS calculation results for other parametrizations of the kagome model \cite{RJ_Kagome} (Fig.~\ref{fig:DOS}). We observe transitions to metallic states, as the ADOS at nodal energies becomes non-vanishing. At much higher disorder, we observe Anderson localization, accompanied by the flattening of DOS (dashed lines). For phases with $(t,t') = (1, 0)$ and $(t,t') = (0, 1)$, the presence of flat band complicates the study of criticality at the quadratic node with $\chi = 1$, therefore a model with $(t,t') = (1, -0.2)$, introducing additional dispersion and making the node accessible, is studied in the main text. For comparison, we note that the qualitative ADOS evolution on adding disorder is similar to graphene, for the energies at which the nodes that can be annihilated ($\pm i$) reside. 

\subsection{Average conductivities evaluated with KPM}

\begin{figure*}
\includegraphics[width=\linewidth]{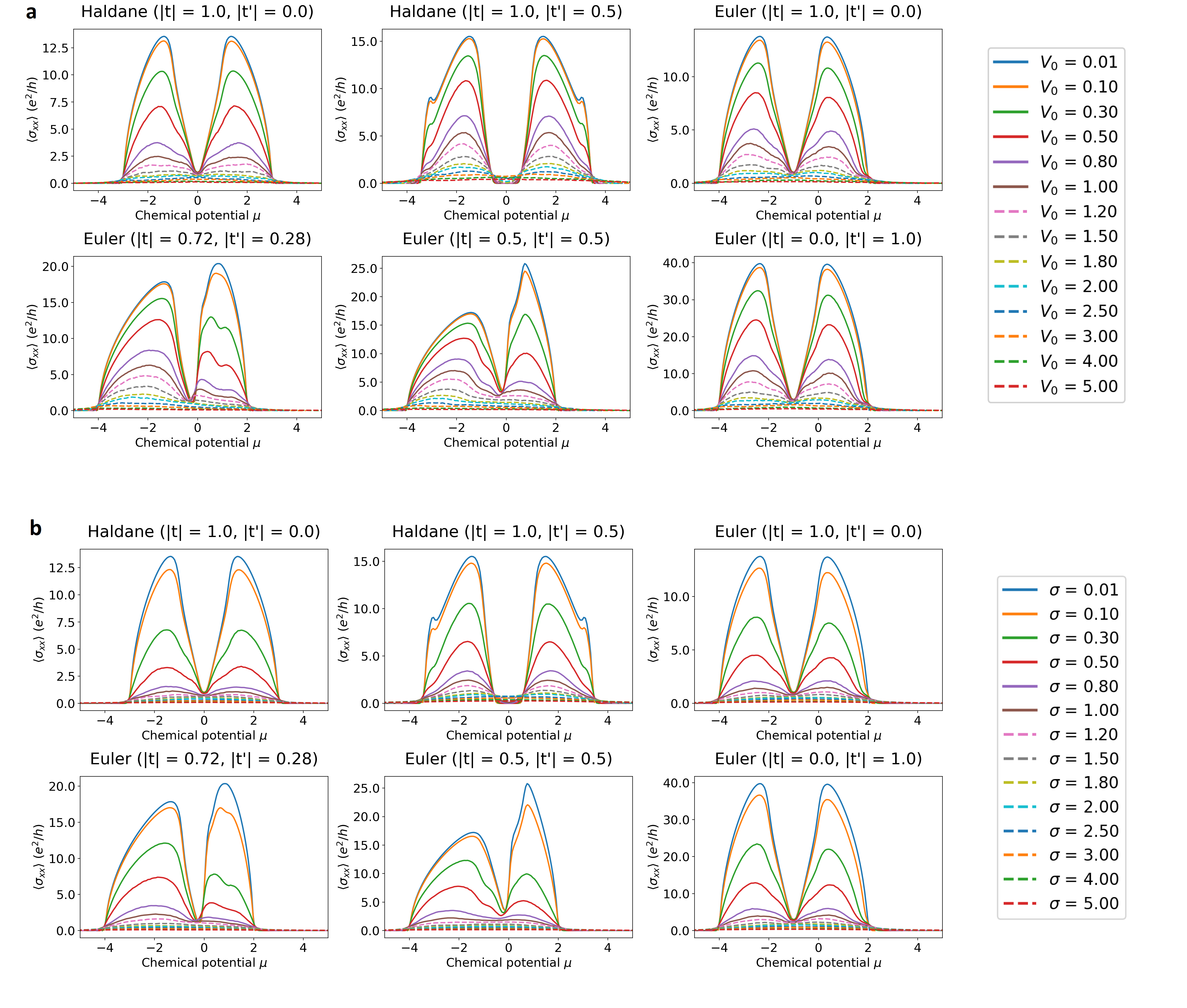}
\caption{Average conductivity of kagome Euler semimetals (and Haldane model, incl. its trivial phase --- graphene with nearest-neighbor hopping $t = 1$, next-nearest-neighbor hopping $t' = 0$, hence vanishing Chern number) subject to  \textbf{(a)}: uniform disorder, \textbf{(b)}: Gaussian disorder. The plots were obtained for systems consisting of $150 \cross 150$ unit cells. From the scaling of conductivity around chemical potentials corresponding to nodes or gap closings, the dynamical scaling exponents ($z$) can be confirmed.}
\label{fig:CondXX}
\end{figure*}

\begin{figure*}
\includegraphics[width=\linewidth]{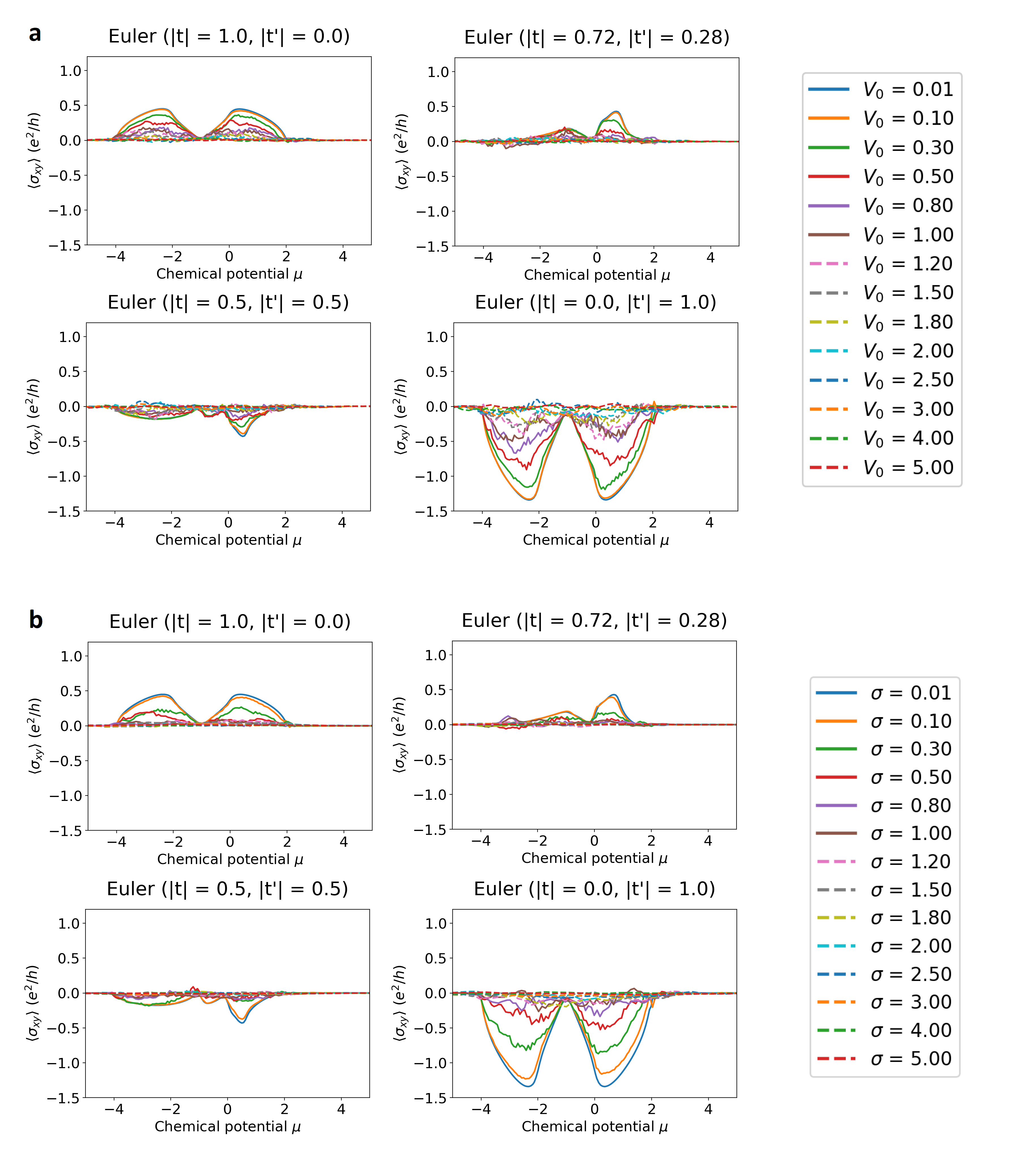}
\centering
\caption{Anomalous conductivity in disordered Euler semimetal models on a kagome lattice. \textbf{(a)}: Uniform disorder. \textbf{(b)}: Gaussian disorder. We observe the conductivity to decay analogously for both types of onsite disorder. The negative values reflect the antisymmetry of the conductivity tensor $\sigma_{xy} = -\sigma_{yx}$, while the plots were obtained for systems consisting of $150 \cross 150$ unit cells. The anomalous conductivity corresponding to the presence of edge states found in Euler Hamiltonians is generically not quantized.}
\label{fig:Kagome_xy}
\end{figure*}

\begin{figure*}
\includegraphics[width=\linewidth]{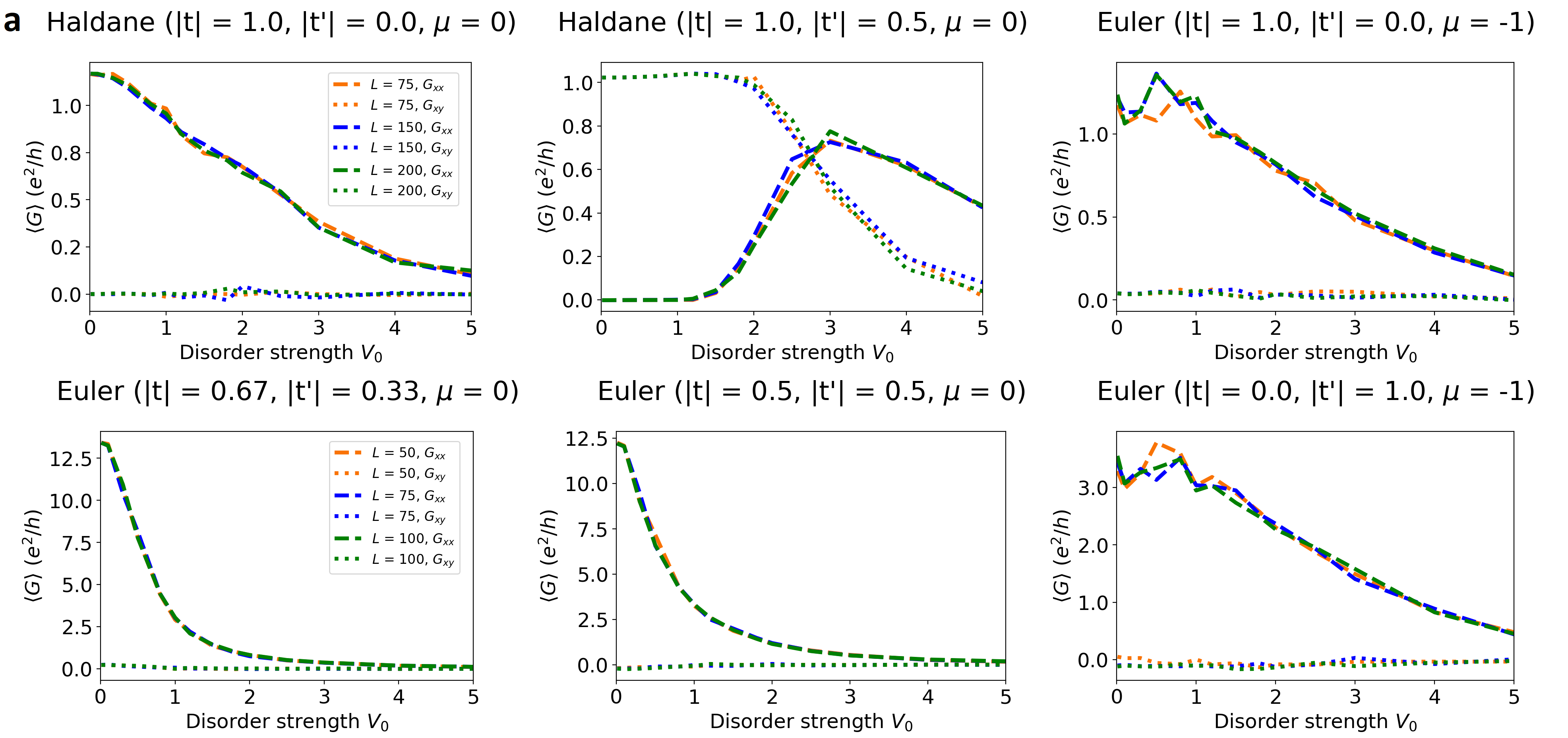}
\includegraphics[width=\linewidth]{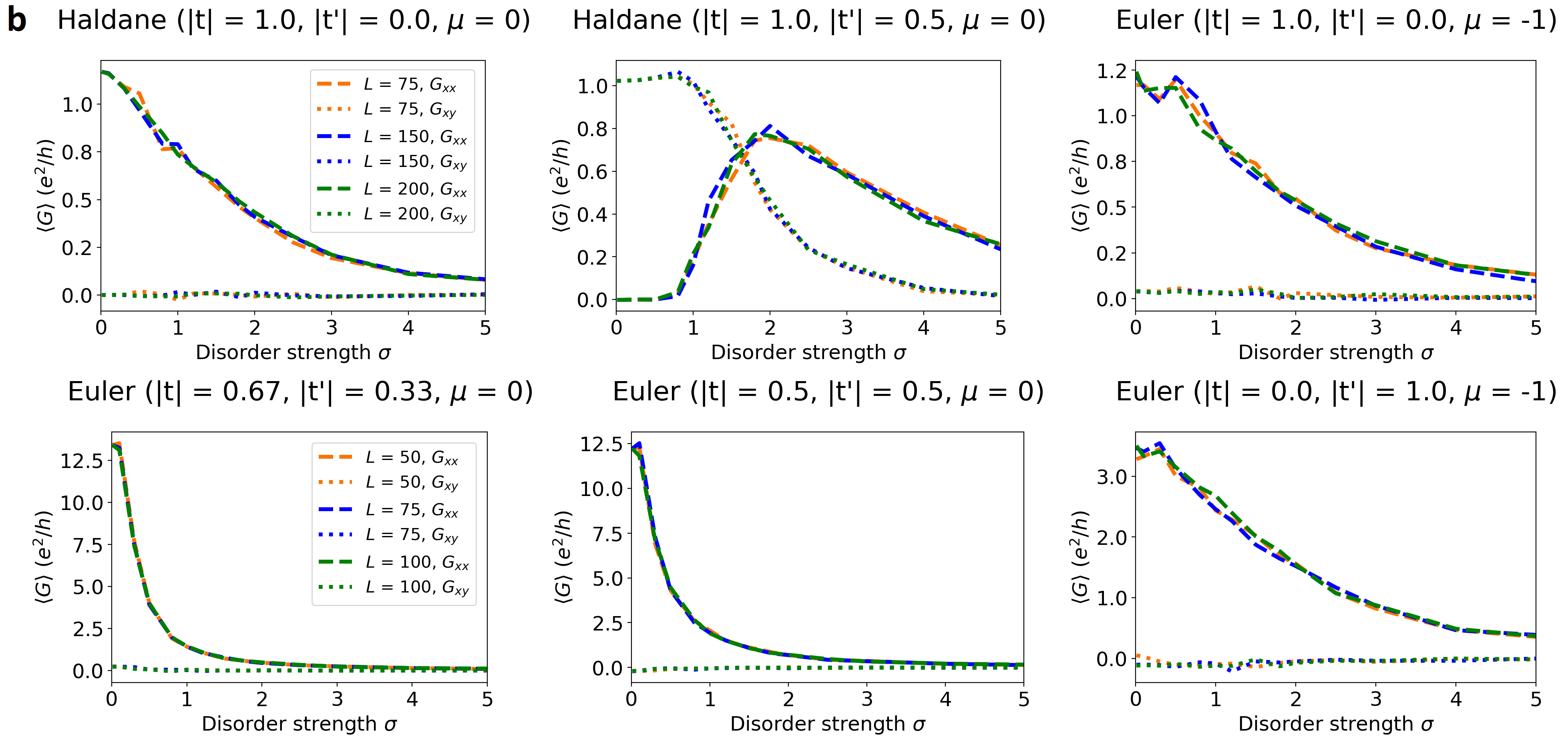}
\caption{Scaling of
conductance in kagome Euler semimetals (and graphene/Haldane model).
\textbf{(a)}: On adding uniform disorder.
\textbf{(b)}: On adding Gaussian disorder. The conductance $G_{ij} = \sigma_{ij} L^{(d-2)}$ is size-independent and equal to
conductivity for two-dimensional systems. We observe that the results for Gaussian and uniform onsite disorders are qualitatively consistent, while both phases with nodes at energy not overlapping with the dispersion of any band pockets, i.e. $(t, t') = (1, 0)$ and $(t, t') = (0, 1)$,  exhibit an intermediary conductance 
behavior corresponding to the development of a metallic phase.}
\label{fig:ScalingCond}
\end{figure*}

Here, the longitudal (Fig.~\ref{fig:CondXX}) and anomalous (Fig.~\ref{fig:Kagome_xy}) conductivity KPM data for the kagome models \cite{RJ_Kagome} subject to disorder is included. Additionally, we show the scaling of conductance with disorder in the kagome semimetals (Fig.~\ref{fig:ScalingCond}). On adding disorder, the conductivities are found to decay at chemical potentials corresponding to the metallic phases. However, at the chemical potential corresponding to the nodes, $\mu = -1$ for $(t,t') = (1, 0)$ and $(t,t') = (0, 1)$, we observe an extended regime up to $W = 1.0 \pm 0.1$, for uniform disorder, which we argue to correspond to the unbraided and delocalized metallic phases, as the QPTs occur (Fig.~\ref{fig:ScalingCond}). We observe that such behavior is not present in the case of graphene.

\subsection{Scaling analysis at QPTs}

\begin{figure*}[t]
\includegraphics[width=\linewidth]{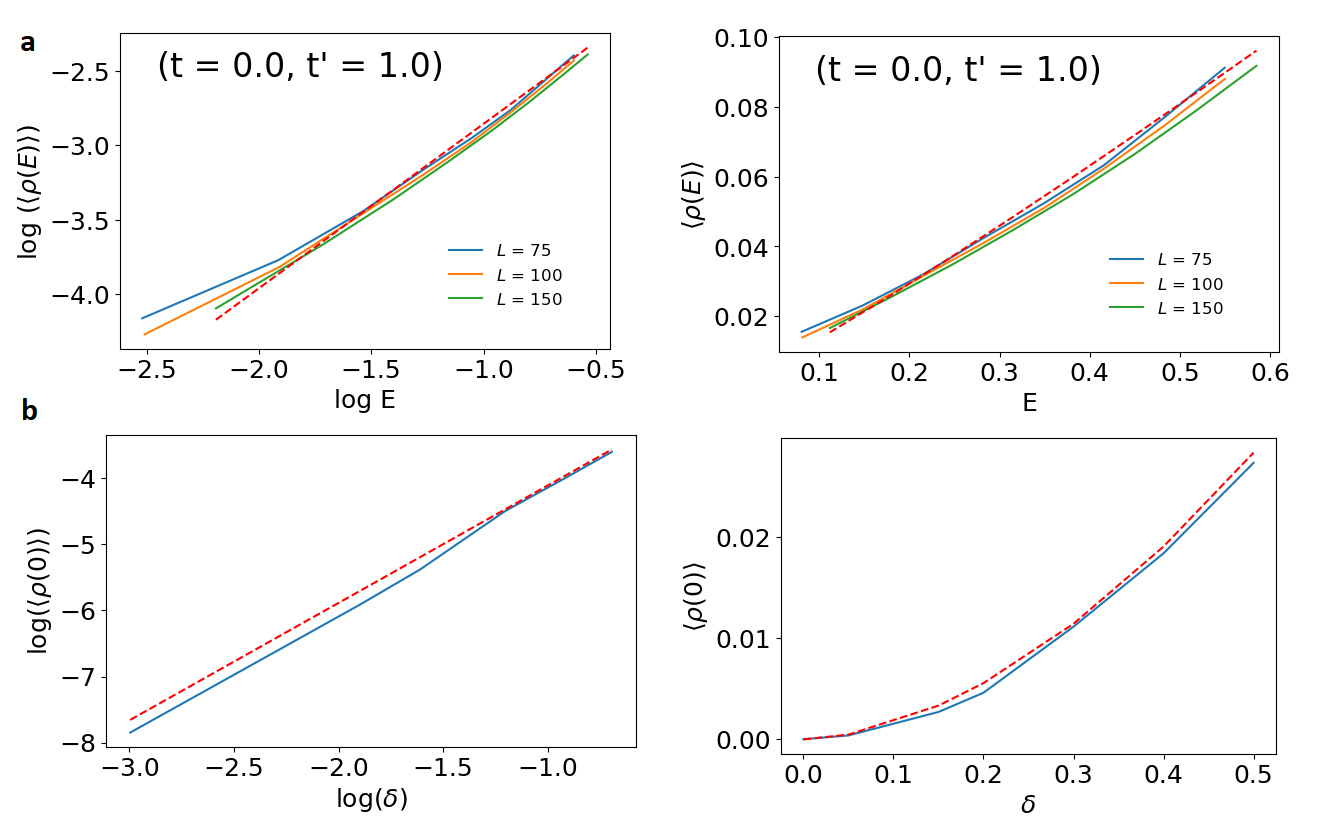}
\caption{ \textbf{(a)}: Scaling fitting to deduce the dynamical scaling exponent $z$ in an unstable semimetal (see main text). The dashed red line represents the fit to different system sizes $(L = 75, 100, 150)$, yielding $z = 0.9 \pm 0.1$ for the oppositely charged nodes. \textbf{(b)}: Fitting for finding the localization length exponent $\nu$ at the gap-closing transition of the protected $\chi = 1$ model system of size $L \cross L = 100 \cross 100$. The dashed line represents the power-law fit, while the blue line represents the scaling of $\rho(0)$, obtaining $\nu = 1.4 \pm 0.1$, as for the other Euler semimetal models protected by a neighboring gap.}
\label{fig:lines}
\end{figure*}

In this section, we show the plots (Fig.~\ref{fig:lines}) from which the scaling exponents, indicated in the scaling analysis section, can be obtained. Firstly, the dynamical exponent $z$ is deduced at the critical disorder ($W_c$) from $\log \rho(E)$ extracted from ADOS vs. $\log E$ around the energies where the nodes reside, or where the gap-closing occurs. Next, using the $z$ deduced, $\log \rho(0)$, with $E = 0$ corresponding to the nodes or gap closing is plotted against $\log \delta$, the reduced disorder $\delta = (W-W_c)/W_c$ above the critical value. From fitting the lines to the log-log plots, prefactors of the scaling power laws are found, and the fit is re-examined in the power law plots.  

\clearpage

\end{document}